\documentclass[3p,twocolumn]{elsarticle}
\usepackage{color}
\usepackage{amssymb}
\journal{Astroparticle Physics}
\begin{document}
\begin{frontmatter}


\title{A fundamental theory based on the Monte Carlo \textit{Time Sequential Procedure}  for the range fluctuations of high energy muons}
\author[1]{Y.Okumura}
\author[1]{N.Takahashi}
\author[2,3]{A.Misaki}
\address[1]{Graduate School of Science and Technology Hirosaki University, Hirosaki, 036-8561, Japan}
\address[2]{Innovative Research Organization, Saitama University, Saitama, 338-8570, Japan}
\address[3]{The Institute for China-Japan Culture Study, Mitaka, Tokyo, 113-0004, Japan}

\begin{abstract}
Lipari and Stanev developed a method for range fluctuation of high energy muons, stressing the importance of accounting for the fluctuations of the energy loss in radiative processes in 1991 and, now, their method has become the basement for the energy determination of high energy muons through the measurement of the Cherenkov light yields due to those muons in KM3 physics. 
Once, Takahashi et al. developed a method for the investigation on the depth intensity relation of high energy muons in which all the stochastic processes concerned are taken into account exactly (1983). Now, we make the method by Takahashi et al. revival for the same purpose of the application to the analysis of future KM3 physics. 
In the present paper, our concern is restricted to the introduction to the fundamental of our method and some subsequent results thereby in which the real simulated behaviors of high energy muons from $10^{12}$eV to $10^{18}$eV, the survival probabilities of high energy and so on are included. The discussion around the practical application of our method to the KM3 physics is entrusted in the subsequent papers. 
As far as the survival probability of high energy muons is concerned, our method gives nearly the same results to Lipari and Stanev's in some regions and gives the deviated results from theirs in another ones. Thus, we examine the application limit of their method and clarify the reason why, comparing with our method. 
The most distinct difference in the both methods may become apparent in the treatment on the Cherenkov light yields spectrum by which one may estimate the energies of the muons concerned. We will mention to them in subsequent papers. 
\end{abstract}

\end{frontmatter}

%
\section{Introduction}
\label{intro}
The studies on depth intensity problem of high energy muons have been one of the most important subjects in traditional cosmic ray physics, relating to elucidation on unknown character of high energy cosmic ray muons, and still have never been lost its importance. Menon and Ramana Murthy~\cite{Menon} wrote an excellent review on this subject at greater depths which describes from the first experiment carried out Simizu tunnel in Japan (1940-1945)~\cite{Miyaz} to the last experiment related to the neutrino carried out in Kolar Gold Mine, India (Up to 1964)~\cite{Miyake1}.  Bugaev et al.~\cite{Buga} have discussed this problem related to the charm production mechanism, adding new data, from, such as, DUMAND, Baikal, MACRO, LVD, NESTOR and others up to 1977. Now, the gigantic projects for neutrino astrophysics called as KM3 detectors are now being developed in the lake, Antarctic and Ocean. In the analysis of these KM3 detectors, the depth intensity relation for high energy muon is utilized for the confirmation of their experimental reliabilities related to other experiments different depths
\footnote{For examples, 
http://baikalweb.jinr.ru/, \\
http://icecube.wisc.edu/, \
http://antares.in2p3.fr/ }.

The theories of range fluctuation of high energy muons are indispensable means in the analysis of the depth intensity relation of high energy muons at certain depths. The theories of range fluctuation are studied in three different manners. The first one is analytical manner~\cite{Mand1}-~\cite{Naumov}, the second one is numerical one~\cite{Oda} and the third one is the Monte Carlo manner~\cite{Bolin}-~\cite{Chirkin}.

At the same time, the theories of range fluctuation of high energy muons offer essential tools for energy determination of high energy muon events from neutrino interactions in KM3 detectors. Their energies, in particular higher energies, are estimated from Cherenkov light signals from muon induced electromagnetic cascade showers, not rather than from muons themselves, but, in spite of this situation, examination of behaviors of muons themselves are essentially important, because they are origins of electromagnetic cascade showers from different modes of interactions due to muons.
   
At present time, studies on the fluctuation of high energy muons have been made by the Monte Carlo method, using electronic computers with great performance, because only this method can clarify fluctuation characters of muons correctly, while the analytical method and numerical one provide essentially their average behavior. The detailed studies around fluctuation of high energy muon events are inevitable, owing to small number of physical events concerned in addition to sharp steepness of the parent neutrino energy spectrum which are the origin of fluctuation. The studies on the range fluctuation of high energy muons by Monte Carlo method had been made even before appearance of electronic computer with great performance~\cite{Bolin}-~\cite{Miyake2}. However, then, they were forced to put more simplified, even more artificially assumptions on their stochastic processes concerned for saving both man powers and computer ones at the period for computation.
   
In 1983 to 84, Takahashi et al.~\cite{Taka1,Taka2} had developed a new Monte Carlo technique where every stochastic process for high energy muon concerned is treated exactly from the stochastic point of view. Namely, the interaction points and the energy division due to interactions concerned (bremsstrahlung, direct electron-positron pair production and photonuclear interaction) are exactly treated in stochastic manner. In the present paper, we call it tentatively \textit{the Time Sequential Procedure}.
   
In 1991, Lipari and Stanev~\cite{Lipari} developed another technique, from a point of the philosophy of Monte Carlo method. They put the diffusion equation on the fluctuation in the form of differential-integral equation and treat it by Monte Carlo technique. They divided the part which is the origin of fluctuation into two parts, namely, the "hard" part (radiation loss part) and the "soft" part ("continuous" energy loss part). In radiation loss part only, they treat fluctuation in Monte Carlo way, but in the "soft" part they deal with the part as "continuous" energy loss. This technique has been adopted by subsequent authors~\cite{Antoni}-~\cite{Chirkin}.  In the present paper, we call it tentatively \textit{the} $V_{cut}$\textit{ Procedure}. Now, \textit{the} $V_{cut}$\textit{ Procedure}~\cite{Lipari}-~\cite{Chirkin} has been extensively utilized in the analysis of muon neutrino events in KM3 detectors [for example, footnote 1].
   
However, in our opinion, taking into account of the fact that Cherenkov light signals due to high energy muons mostly come from the muon induced electromagnetic cascade showers whose origin is either bremsstrahlung or direct electron-positron pair production or photonuclear interaction than from muons themselves in KM3 detector, the energy determinations of high energy muon events inevitably include more ambiguity in the case of \textit{the} $V_{cut}$\textit{ Procedure}, compared with the case of \textit{the Time Sequential Procedure} (see, the section 3 and 4 Conclusion and Outlook ).   

In the present paper, we try to revitalize \textit{the Time Sequential Procedure} in 1983 for the more accurate analysis of high energy muon events in KM3 detector, comparing to the results obtained by \textit{the} $V_{cut}$\textit{ Procedure} which has been well distributed. Here, we restrict our concern to the fundamental and its application will be reported in subsequent papers.

In the present paper, we propose a new method for more accurate calculation on the range fluctuation of high energy muons and conjecture the possible application of this method to the measurements on Cherenkov light in KM3 physics, but we never propose any kind of the code for the practical applications, which is out of the scope of the present paper and it will be discussed in subsequent papers, if necessary.

\section{Fundamental Structure of \textit{the Time Sequential Procedure} and its Validity}
\label{sec:2}
Here, in order to clarify characteristics of \textit{the Time Sequential Procedure} in contrast to \textit{the} $V_{cut}$\textit{ Procedure}, we reproduce our procedure which had already been published in 1984 (in Japanese)~\cite{Taka2}.

\subsection{The mean free paths for stochastic processes and their resultant mean free path}
\label{sec:2.1}
Behaviors of high energy muons are stochastically determined from the elementary processes of bremsstrahlung~\cite{Kelner}, direct electron-positron pair production~\cite{Kokoulin} and photonuclear interaction processes~\cite{Borg}. 
Further information on the cross sections concerned is found in \cite{Groom}.

We treat these processes as stochastic ones as exactly as possible, without introducing any approximation in the energy region in which we are interested
\footnote{
We adopt, $1$ GeV, the minimum energy of the muon for simulation throughout the present paper. The numerical value of $1$ GeV is adopted for the same purpose in \cite{Lipari}.
}. 
In our procedure, these stochastic processes are prepared as independent ones and, therefore, they are easily replaced by the most advanced ones, if necessary, keeping exactness of our logical structure.

Let us denote, differential cross sections for bremsstrahlung, direct electron-positron pair production, and photonuclear interaction, $\sigma_{b}\left(E,E_{b}\right)dE_{b}$, $\sigma_{d}\left(E,E_{d}\right)dE_{d}$ and $\sigma_{n}\left(E,E_{n}\right)dE_{n}$, respectively. Here, $E$ denotes the energy of muon concerned, $E_{b}$, the energy of photon due to bremsstrahlung, $E_{d}$, the energy of electron pair due to direct electron-positron pair production, $E_{n}$, the energy of the hadronic part due to photonuclear interaction, respectively. Then, the mean free paths for different stochastic processes are energy dependent of the muon concerned and they are given as follows:
\\
For bremsstrahlung processes,
\begin{equation}        
\lambda_{b}\left(E\right)= \frac{1}{\frac{N}{A}\int_{E_{b,min}}^{E_{b,max}} \sigma_{b}\left(E,E_{b}\right)dE_{b}}\\
\end{equation}
Here, $E_{b,min}$, the lower limit of the integral of Eq.(1), the minimum energy for the emitted photons, is taken $1.02$ MeV which denotes the minimum energy for electron pair production by photon. 

The integrations for direct electron-positron pair production and photonuclear interaction are performed over kinematically allowable ranges.
\\
For direct electron-positron pair production processes,
\begin{equation}        
\lambda_{d}\left(E\right)= \frac{1}{\frac{N}{A}\int_{E_{d,min}}^{E_{d,max}} \sigma_{d}\left(E,E_{d}\right)dE_{d}}\\
\end{equation}
For photonuclear interaction processes,
\begin{equation}        
\lambda_{n}\left(E\right)= \frac{1}{\frac{N}{A}\int_{E_{n,min}}^{E_{n,max}} \sigma_{n}\left(E,E_{n}\right)dE_{n}}\\
\end{equation}
, where $N$ and $A$ denote Avogadro number and atomic mass number, respectively.
Similarly, $E_{d,min}/E$ and $E_{n,min}/E$ are always chosen in such a way that the differential cross sections concerned are expressed exactly above $E_{min}$.

Also, $\lambda_{total}\left(E\right)$, the resultant mean free path for these stochastic processes are given as,
\begin{equation}        
\frac{1}{\lambda_{total}\left(E\right)}= \frac{1}{\lambda_{b}\left(E\right)}+\frac{1}{\lambda_{d}\left(E\right)}+\frac{1}{\lambda_{n}\left(E\right)}
\end{equation}

\subsection{Determination of the kind of the stochastic process and the real free path for the stochastic process concerned}
\label{sec:2.2}
By using Eq.(1) to (4), we can determine the interaction points of muons for different stochastic processes in the following. The first, for the purpose, lets us define
$\xi_{b}\left(E\right)$ and $\xi_{d}\left(E\right)$ as follows;
\begin{equation}        
\xi_{b}\left(E\right)=\frac{1/\lambda_{b}\left(E\right)}{1/\lambda_{total}\left(E\right)}
\end{equation}
\begin{equation}        
\xi_{d}\left(E\right)=\frac{1/\lambda_{b}\left(E\right)+1/\lambda_{d}\left(E\right)}{1/\lambda_{total}\left(E\right)}
\end{equation}
The second, we sample randomly $\xi_{1}$, a uniform random number between (0,1). If $\xi_{1} \le \xi_{b}\left(E\right)$, then we recognize the interaction occurs due to bremsstrahlung. If $\xi_{b}\left(E\right) < \xi_{1} \le \xi_{d}\left(E\right)$, then, we understand the direct electron-positron pair production occurs. If $\xi_{1} > \xi_{d}\left(E\right)$, then, we understand that photonuclear interaction occurs.

Again, we sample a new $\xi_{2}$, randomly from uniform random number between (0,1). Then, we can determine the interaction points $\Delta t\left(E\right)$ for the specified stochastic processes according the following criterion. In the case of the occurrence of bremsstrahlung processes ( for $\xi_{1} \le \xi_{b}\left(E\right)$ ),
\begin{equation}        
\Delta t_{b}\left(E\right) = -\lambda_{total}\left(E\right)log\xi_{2}
\end{equation}
In the case of the occurrence of direct electron-positron pair production processes (for  $\xi_{b}\left(E\right) < \xi_{1} \le \xi_{d}\left(E\right)$ ),
\begin{equation}        
\Delta t_{d}\left(E\right) = -\lambda_{total}\left(E\right)log\xi_{2}
\end{equation}
In the case of the occurrence of photonuclear  interaction processes ( for $\xi_{1} > \xi_{d}\left(E\right)$ )
\begin{equation}        
\Delta t_{n}\left(E\right) = -\lambda_{total}\left(E\right)log\xi_{2}
\end{equation}

\subsection{The effect of both the "continuous" energy losses and the usual ionization loss over the muon propagation in the Time Sequential Procedure.}

We are taken into account of the "continuous" energy loss (the first term (the "soft" part) of Eq.(14)) in addition to the usual ionization loss. The effect of the "continuous" energy loss may be evaluated in two ways. One is Tamura's method \cite{Tamura} in which "continuous" energy loss is treated together with the usual ionization loss. The other is Adachi's method \cite{Adachi} in which bremsstrahlung cross section is deformed so as to neglect the "continuous" energy loss \cite{Adachi}. Then, the effect due to the "soft" part the first term of Eq.(14) is compensated by the increase of the "hard" part due to bremsstrahlung.

Then, we adopt the Tamura's method which is essentially the same as Lipari and Stanev \cite{Lipari}.


In $the Time Sequential Procedure$, $v_{cut}$ is defined in the same way as \textit{the} $V_{cut}$ \textit{Procedure} and $ v_{cut} = E_{b,min}/E$, where $E$ and $E_{b,min}$,  denote the energy of the muon concerned and the minimum energy of the emitted photons due to bremsstrahlung, respectively. Due to the adoption of $1.02$ MeV ($\sim 10^{6}$ eV) as the lower limit of the integrals of Eqs.(1) and (10), the "continuous" energy loss per muon radiation length($\sim 1500$ meter in water) is $1.02$ MeV ($\sim 10^{6}$ eV) by the definition, irrespective of the energies of the muons concerned, which is far smaller compared with the usual ionization loss, $\sim 3\times 10^{11}$ eV per muon radiation length ($\sim 2$ MeV/(g/cm$^{2}$)) by the five order of the magnitude. 
Consequently, we can completely neglect the "continuous" energy loss due to bremsstrahlung even compared with the usual ionization loss, irrespective of the energies of the muons concerned. Furthermore, it should be noticed that we have concern in the behaviors of the muons whose primary energies extend from $10^{9}$ eV ($1$ GeV) to $10^{18}$ eV for KM3 physics. Considering the dimension of the KM3 detector is of one kilometer, we can neglect the usual ionization for the muons above $10^{13}$ eV. However, we consider the usual ionization loss as well as the "continuous" energy loss in our procedure.

Thus, we can state that the energy losses due to muons come essentially from the "radiative" processes (bremsstrahlung, direct electron-positron pair production and photo nuclear interaction) in our procedure. Namely, we can simulate exactly the energies of the emitted photons down to $1$ MeV (due to bremsstrahlung and photo nuclear interaction) as well as the electrons (positrons) (due to direct electron-positron pair production). Thus, these emitted energies of the particles (electrons and photons) induce the photon initiated electromagnetic cascade showers and the electron(positron) initiated electromagnetic cascade showers which are the sources of the Cherenkov light yields. The capability of the complete neglect of the "continuous" energy loss is the characteristics of our \textit{Time Sequential Procedure} by which our algorithm makes it possible to be constructed in consistent manner. 

The main subject of the present paper is to discuss the behavior of the muon and, therefore, the behaviors of the electromagnetic showers as well as those of the subsequent Cherenkov light yields will be examined in the subsequent papers.

\subsection{Determination of the emitted energy loss ($E_{b}$ or $E_{d}$ or $E_{n}$) from the specified stochastic processes}
\label{sec:2.3}
Under the determination of the interaction points due to the specified stochastic processes, by using $\xi_{1}$ and $\xi_{2}$, in the previous subsection, here, the energy losses $E_{b}$ , $E_{d}$ and $E_{n}$ from the specified stochastic processes are given as follows. For sampled $\xi_{3}$ which is obtained randomly from the uniform random number between (0,1), we solve the following equations for respective interactions in order to obtain $E_{b}$ or $E_{d}$ or $E_{n}$. 

For bremsstrahlung process,
\begin{equation}        
\label{xi_b}
\xi_{3} = \frac{\int_{E_{b,min}}^{E_{b}} \sigma_{b}\left(E,E_{b}\right)dE_{b}}{\int_{E_{b,min}}^{E_{b,max}} \sigma_{b}\left(E,E_{b}\right)dE_{b}}
\end{equation}
$E_{b}$, the emitted photon stochastically sampled from Eq.(10), is expected to generate the electromagnetic cascade shower. The reason why the minimum energy is taken $1.02$ MeV is that the minimum of the electromagnetic cascade shower initiated photons is of two electrons from the electron pair production by photons. The energy dissipation below $E_{b,min}$, $1.02$ MeV, is treated in the same way like the soft term (the "continuous" energy loss) in the [$dE/dx$] method (see, the soft term of Eq.(14) in the present paper) 

For direct electron-positron pair production,
\begin{equation}        
\label{xi_d}
\xi_{3} = \frac{\int_{E_{d,min}}^{E_{d}} \sigma_{d}\left(E,E_{d}\right)dE_{d}}{\int_{E_{d,min}}^{E_{d,max}} \sigma_{d}\left(E,E_{d}\right)dE_{d}}
\end{equation}

For photonuclear interaction,
\begin{equation}        
\label{xi_n}
\xi_{3} = \frac{\int_{E_{n,min}}^{E_{n}} \sigma_{n}\left(E,E_{n}\right)dE_{n}}{\int_{E_{n,min}}^{E_{n,max}} \sigma_{n}\left(E,E_{n}\right)dE_{n}}
\end{equation}
In Eq.(10) to (12) for respective interaction, the quantities to be obtained are $E_{b,d,n}$, emitted energy losses for given $E$, energy of the muon concerned. The quantities of $E_{b,d,n}$ are extended to $E_{b,d,n,min}$ to $E_{b.d,n,max}$. For sampled $\xi_{3}$, we can solve these equation numerically and obtain $E_{b,d,n}$ finally. Thus, we can determine $E_{b,d,n}$, the energy losses for the specified stochastic process at determined interaction point. A flow chart for the fundamental structure of \textit{the Time Sequential Procedure} is given Fig.1.

\begin{figure}[h]          
\begin{center}
\resizebox{0.48\textwidth}{!}{\includegraphics{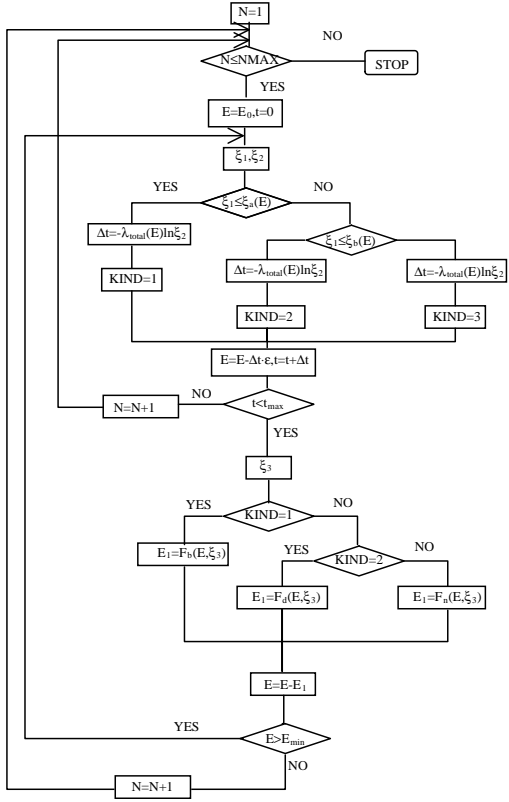}}     
\caption{Flow Chart for the fundamental structure of \textit{the Time Sequential Procedure}.}
\label{Flow} 
\end{center}
\end{figure}

\subsection{On the validity of \textit{the Time Sequential Procedure}}
\label{sec:2.4}
Generally speaking, the verification of the validity of the Monte Carlo method concerned is pretty difficult. 
For the verification of our procedure, it is desirable to compare the physical results obtained by \textit{the Time Sequential Procedure} with the corresponding results obtained by the analytical method which is methodologically independent of the Monte Carlo method concerned.

Once, Misaki and Nishimura~\cite{Misaki1} had developed an analytical theory for range fluctuation of high energy cosmic ray muons based on the Nishimura-Kamata formalism on electron shower theory to apply to study depth intensity relation muon underground. 
The analytical theory could be solved rigorously only under the incident muon energy spectrum whose indices (in integral) $\gamma $= 2, 3, 4, , , . For the comparison of the results obtained by \textit{the Time Sequential Procedure} with those obtained by the analytical theory, Takahashi et al.~\cite{Taka1,Taka2} had calculated the depth dependence of the average energies of muons at various depths under the same incident muon energy spectrum which the analytical theory utilized and compared their results with corresponding ones obtained by the analytical theory.

\begin{figure}[h]           
\begin{center}
\resizebox{0.45\textwidth}{!}{\includegraphics{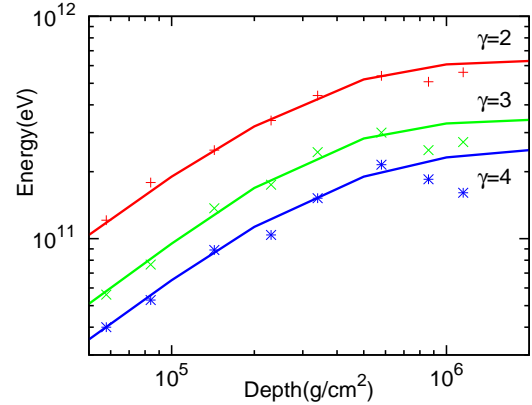}}   
\caption{The average energies of the muons. The lines denote Misaki and Nishimura, while symbols ours.}
\label{fig}
\end{center}
\end{figure}

We reproduce them in Fig.2 from the previous work~\cite{Taka2}. The agreement between them is quite well, taking into account of the difference in both the cross sections concerned and their numerical evaluation method. Namely, we can say the logical structure of \textit{the Time Sequential Procedure} is well established from the point of the validity of Monte Carlo method concerned. It is needless to say that \textit{the Time Sequential Procedure} can be applicable to any incident muon energy spectrum.

\subsection{Results directly derived by \textit{the Time Sequential Procedure}}
\subsubsection{The diversity of individual muon behavior 'Needle' structure of the energy losses from high energy muons}
\label{sec:2.4.1}
In subsections \textit{2.2} and \textit{2.3}, by random sampling procedure, we show how to determine both interaction points for the specified stochastic processes and subsequent their energy losses for the interactions concerned. In the present subsections, we show some examples of the energy losses along the passage of the high energy muons.

\begin{figure*}[!t]
\begin{center}
\resizebox{1.0\textwidth}{!}{\includegraphics{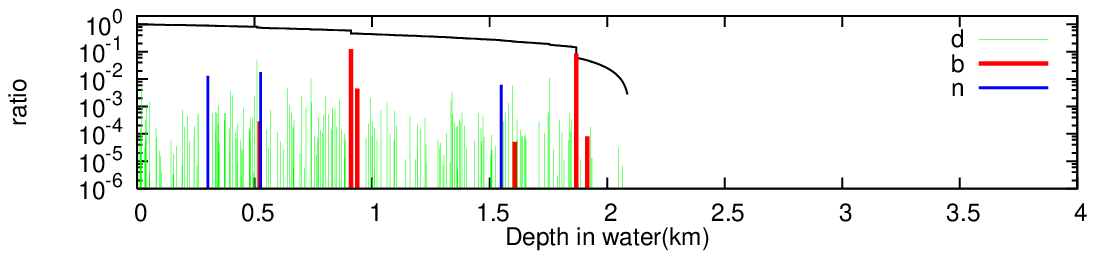}}	      
\caption{The fractional energy loss with \textit{the shortest range} as the function of the depth for $10^{12}$eV muon. A line graph in the upper denotes fractional muon energy as the function of the depth. [\textbf{b}] denotes the fractional energy loss due to bremsstrahlung, [\textbf{d}] due to direct electron-positron pair production and [\textbf{n}] due to photonuclear interaction. [\textbf{b}], [\textbf{d}] and [\textbf{n}] are utilized as the same meaning up to Figure 13.}
\label{fig:ELS12}
\resizebox{1.0\textwidth}{!}{\includegraphics{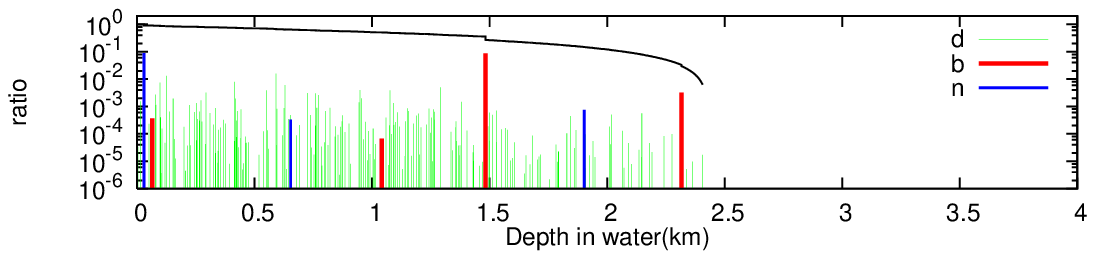}}	      
\caption{The fractional energy loss with \textit{the average-like range} for $10^{12}$eV muon muon together with the fractional energy of the muon.}
\label{fig:ELA12}
\resizebox{1.0\textwidth}{!}{\includegraphics{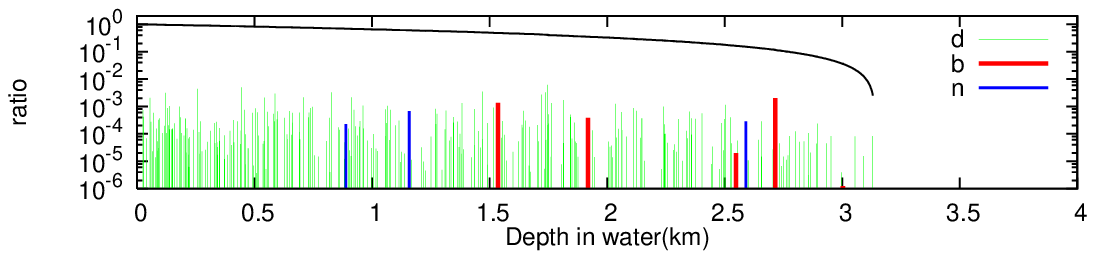}}	      
\caption{The fractional energy loss with \textit{the longest range} for $10^{12}$eV muon muon together with the fractional energy of the muon.}
\label{fig:ELL12}
\end{center}
\end{figure*}

In Fig.3 to Fig.13, we show three different categories of the 'Needle' structure of the energy losses for the same primary energies. 'Needles' denote fractional energy losses due to a specified stochastic processed such as, bremsstrahlung, direct electron-positron pair production or photonuclear interaction at respective interaction points. 
We illustrate several typical structures of the energy loss of muons which start, having primary energy $E_{0}$ and reaching $E_{min}$, namely the behaviors of the muons with \textit{the shortest range}, with \textit{the longest range} and with \textit{the average-like range} for the same primary energy and the same starting point. 
\textit{The shortest range} denotes the muon with the shortest range among all sampled muons, while \textit{the longest range} does the muon with the longest range among all sampled muons, and \textit{the average-like range} does the muon with the range whose is the nearest to the averaged range among all sampled muons. Total sampling numbers per respective primary energy are 100,000.

In these figures, we can recognize the diversities of muon behaviors for the same primary muon energies with regard to their ranges (or their energy losses). All interaction points due to the processes of bremsstrahlung, direct electron-positron pair production and photonuclear interaction and all energy losses due to these elementary processes at respective points due to these processes are recorded. In order to clarify the diversities among the real range distributions (or real energy loss distributions), we examine the muons with \textit{the shortest range}, the muons with \textit{the longest range} and the muons with \textit{the average-like range} in more detail.

\begin{table*}[!t]              
\begin{center}
\caption{The details of the characteristics on the muons with \textit{the shortest range}, \textit{the average-like range},\textit{the longest range} and \textit{the average range}.}
\label{tab:ELT}
\scalebox{0.9}[0.85]{
\begin{tabular}{c|c|c|c|c|c|c|c}
\hline
\noalign{\smallskip}
	              & Range   & Energy loss  & Number of   & Energy loss by  & Number of   & Energy loss    & Number of \\
$E_{0}=10^{12}eV$ &	[km]	& by brems[eV] & interaction & direct pair[eV] & interaction & by nuclear[eV] & interaction \\
\noalign{\smallskip}
\hline
$<$Average$>$  &  2.43     &  1.10$\times 10^{11}$ & 4.74     & 1.57$\times 10^{11}$ & 243                 & 4.54$\times 10^{10}$ & 3.44\\ \hline
Average-like   &  2.43     &  8.97$\times 10^{10}$ & 4        & 1.34$\times 10^{11}$ & 221                 & 8.86$\times 10^{10}$ & 3 \\ \hline
Shortest       &  2.09     &  2.15$\times 10^{11}$ & 6        & 1.52$\times 10^{11}$ & 208                 & 3.72$\times 10^{10}$ & 3 \\ \hline
Longest        &  3.14     &  3.80$\times 10^{9}$  & 5        & 1.04$\times 10^{11}$ & 299                 & 1.19$\times 10^{9}$  & 3 \\ \hline
\noalign{\smallskip}
$E_{0}=10^{15}eV$\\ 
\noalign{\smallskip}
\hline
$<$Average$>$  & 1.78$\times 10^{1}$  & 3.53$\times 10^{14}$ & 48.1      & 4.74$\times 10^{14}$ & 6.80$\times 10^{3}$ & 1.67$\times 10^{14}$ &  5.50$\times 10^{1}$\\ \hline
Average-like   & 1.78$\times 10^{1}$  & 7.50$\times 10^{14}$ & 49        & 2.35$\times 10^{14}$ &  5489               & 9.31$\times 10^{12}$ &  37\\ \hline
Shortest       & 9.44$\times 10^{-1}$ & 8.66$\times 10^{14}$ &  2        & 1.34$\times 10^{14}$ &   367               & 5.90$\times 10^{10}$ &   1\\ \hline
Longest        & 3.50$\times 10^{1}$  & 7.53$\times 10^{13}$ & 71        & 8.02$\times 10^{14}$ & 13722               & 1.11$\times 10^{14}$ & 105\\ \hline
\noalign{\smallskip}
$E_{0}=10^{18}eV$\\ 
\noalign{\smallskip}
\hline
$<$Average$>$ & 3.28$\times 10^{1}$ & 3.37$\times 10^{17}$ & 1.08$\times 10^{2}$ & 4.39$\times 10^{17}$ & 2.57$\times 10^{4}$ & 2.25$\times 10^{17}$ & 1.72$\times 10^{2}$\\ \hline
Average-like  & 3.28$\times 10^{1}$ & 1.68$\times 10^{17}$ & 118         & 5.58$\times 10^{17}$ & 29321               & 2.74$\times 10^{17}$ & 196\\ \hline
Shortest      & 7.72$\times 10^{0}$ & 8.75$\times 10^{17}$ &  28         & 1.19$\times 10^{17}$ &  5760               & 6.23$\times 10^{15}$ &  40\\ \hline
Longest       & 5.78$\times 10^{1}$ & 5.71$\times 10^{16}$ & 162         & 5.77$\times 10^{17}$ & 46542               & 3.66$\times 10^{17}$ & 277\\ \hline
\noalign{\smallskip}
\end{tabular}
}
\end{center}
\end{table*}

In Table 1, we show numerically the characteristics of an individual muon with \textit{the shortest range}, \textit{the average-like range}, \textit{the longest range}, in addition to their average range for $10^{12}$eV, $10^{15}$eV and $10^{18}$eV.

In Fig.3 to Fig.5, we give the characteristic behaviors with \textit{the shortest range}, \textit{the average-like range} and \textit{the longest range}, that is, their energy loss for the specified interaction as the function of the depth traversed for the primary energy of $10^{12}$eV in water
\footnote{In order to understand the situation visually the characteristic behaviors of high energy muons which are shown Fig.3 to Fig.13, we suggest the readers to look at the pictures with colors in the WEB page.}
. 
In these figures, we utilize the same scale in depth to clarify the diverse behaviors by the same incident energies, namely, those with \textit{the shortest range}, with \textit{the average-like range} and with \textit{the longest range}, respectively. In figures, the abscissa denotes the depths where the specified interactions occur. The 'needles' (expressed in ordinate) with different colors at different depths denote ratios of the energy losses due to direct electron-positron pair production (green, \textbf{d}), bremsstrahlung (red, \textbf{b}) and photonuclear interaction (blue, \textbf{n}) to their primary energy, respectively. The abrupt changes in them are due to the catastrophic energy losses for muons (see, footnote 3). It is easily understood that one sees the fluctuation effect rather weak in the energy of $10^{12}$eV.

It is seen from figures and Table 1 that there is not so big difference between the case with \textit{the shortest range} and one with \textit{the longest range} for $10^{12}$eV. In the case with \textit{the shortest range} (Fig.3), we find two catastrophic energy losses (at $\sim 910$ meters and $\sim 1870$ meters) due to two bremsstrahlungs play the important role in the range. In the case with \textit{the average-like range} (Fig.4), we can find one catastrophic energy loss due to bremsstrahlung at $\sim 1.48$ kilometer. However, in the case of \textit{the longest range }(Fig.5), we cannot find the catastrophic energy losses due to bremsstrahlung and, instead, we can find that almost energy losses are due to many number($\sim 300$) of direct electron-positron pair production events.

\begin{figure*}[!t]
\begin{center}
\resizebox{1.0\textwidth}{!}{\includegraphics{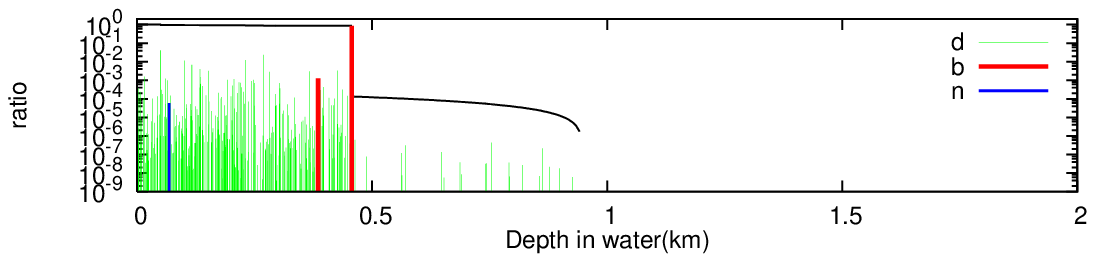}}	  
\caption{The fractional energy loss with \textit{the shortest range} for $10^{15}$eV muon together with the fractional energy of the muon. The figure is a magnification of Figure 7.}
\label{fig:ELS15_2km}
\resizebox{1.0\textwidth}{!}{\includegraphics{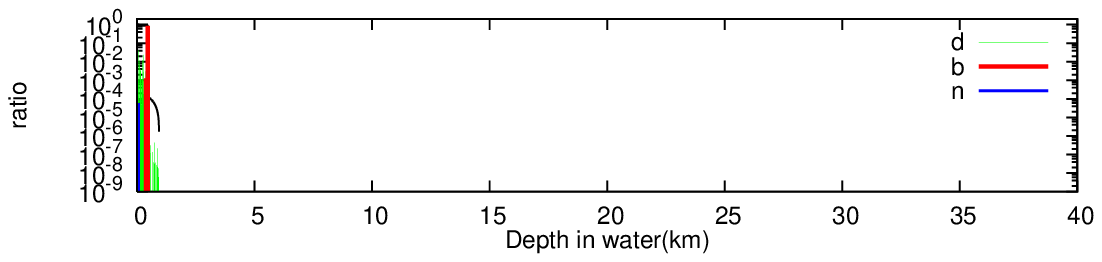}}	      
\caption{The fractional energy loss with \textit{the shortest range} for $10^{15}$eV muon together with the fractional energy of the muon.}
\label{fig:ELS15}
\resizebox{1.0\textwidth}{!}{\includegraphics{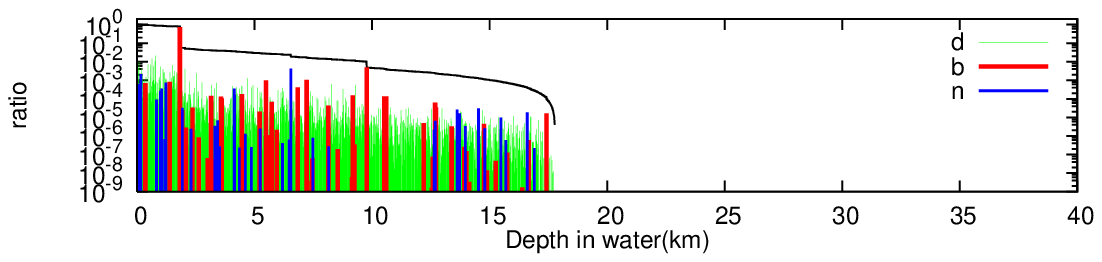}}	      
\caption{The fractional energy loss with \textit{the average-like range} for $10^{15}$eV muon together with the muon energy.}
\label{fig:ELA15}
\resizebox{1.0\textwidth}{!}{\includegraphics{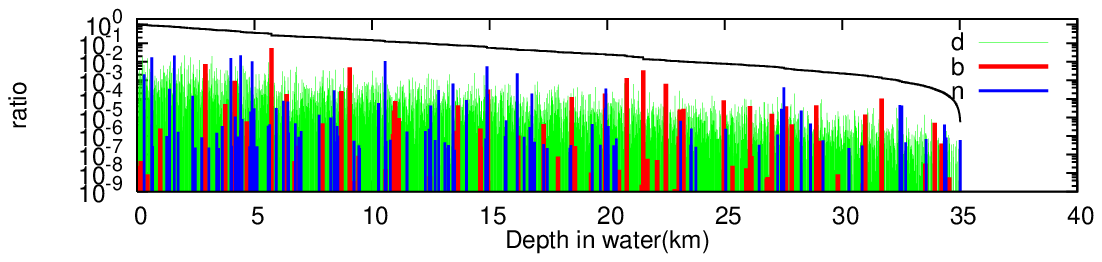}}	      
\caption{The fractional energy loss with \textit{the longest range} for $10^{15}$eV muon together with the fractional energy of the muon.}
\label{fig:ELL15}
\end{center}
\end{figure*}

In Fig.7 to Fig.9, we give the typical diversities for primary energy of $10^{15}$eV similarly for primary energy of $10^{12}$eV. In these figures, the diversities for \textit{the shortest range}, \textit{the average-like range} and \textit{the longest range} are compared explicitly expressed in the same scale. Fig.6 shows the same in Fig.7 in extended scale. Combined with Table 1, \textit{the shortest range}, $\sim 940$ meter (Fig.6), is far shorter compared with \textit{the longest range}, $\sim 35.0$ kilometers (Fig.9). It is seen from Fig.6 and the Table 1 that bremsstrahlung plays a decisive role as the cause of catastrophic energy loss in the case of \textit{the shortest range}, ( $\sim 96.5$\% of total energy up to$ \sim 450$ meters). $86.6$\% of the total energy is lost by $2$ bremsstrahlungs, $13.4$\% by $367$ direct electron-positron pair productions and $5.9 \times10^{-3}$\% by $1$ photonuclear interaction. In Fig.9, we give the case for \textit{the longest range}. Here, large numbers of direct electron-positron pair production with rather small energy loss play an important role, as shown similarly in Fig.5. Here, $80.2$\% of the total energy is lost by $13722$ direct electron-positron pair productions, $7.53$\% by $71$ bremsstrahlungs and $11.1$\% by $105$ photonuclear interactions. In Fig.8, combined with Table 1, we give the case with \textit{the average-like range}.  Here, $23.5$\% of the total energy is lost by $5489$ direct electron-positron pair productions, $75.0$\% by $49$ bremsstrahlungs and $0.93$\% by $37$ photonuclear interactions, while in the real averages ($100,000$ samples), $47.4$\% of the total energy is lost by $6800$ direct electron-positron pair productions, $35.3$\% by $48.1$ bremsstrahlungs and $16.7$\% by $55.0$ photonuclear interactions.

\begin{figure*}[!t]
\begin{center}
\resizebox{1.0\textwidth}{!}{\includegraphics{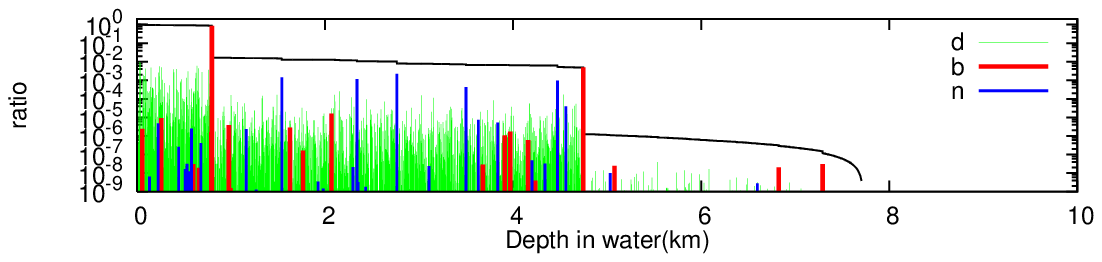}}    
\caption{The fractional energy loss with \textit{the shortest range} for $10^{18}$eV together with fractional energy of the muon. The figure is a magnification of Figure 11.}
\label{fig:ELS18_10km}
\resizebox{1.0\textwidth}{!}{\includegraphics{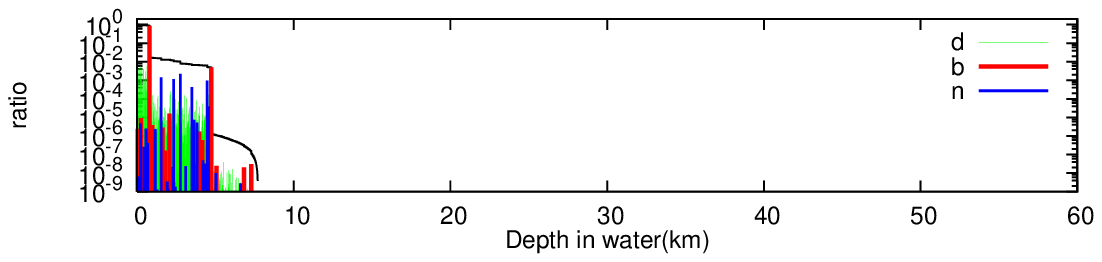}}	      
\caption{The fractional energy loss with \textit{the shortest range} for $10^{18}$eV muon together with the fractional energy of the muon.}
\label{fig:ELS18}
\resizebox{1.0\textwidth}{!}{\includegraphics{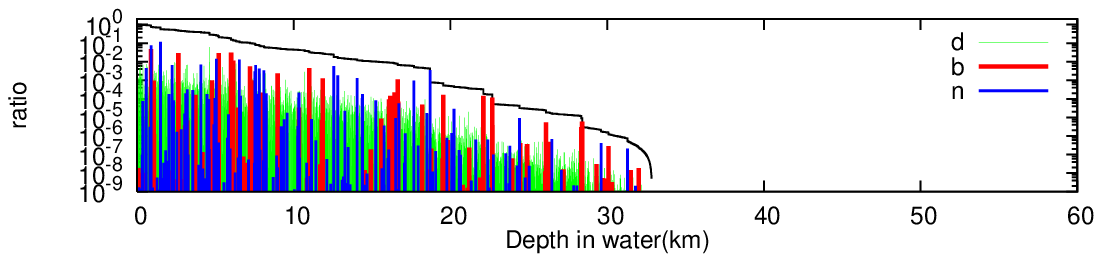}}	      
\caption{The fractional energy loss with \textit{the average-like range} for $10^{18}$eV muon together with the fractional energy of the muon. }
\label{fig:ELA18}
\resizebox{1.0\textwidth}{!}{\includegraphics{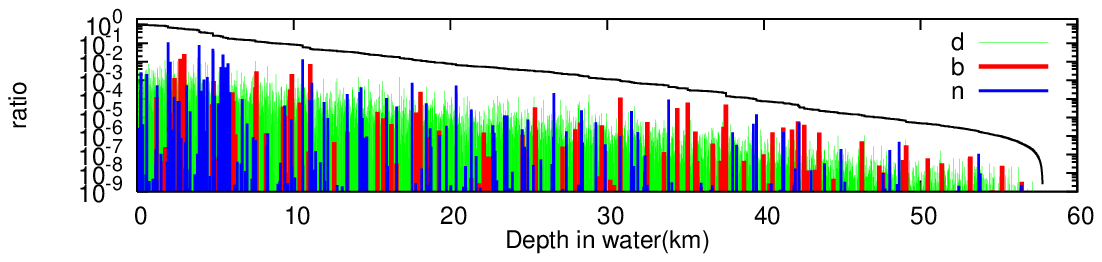}}	      
\caption{The fractional energy loss with \textit{the longest range} for $10^{18}$eV muon together with the fractional energy of the muon.}
\label{fig:ELL18}
\end{center}
\end{figure*}

In Fig.10 to Fig.13, combined with Table 1, we show the similar relations for $10^{18}$eV muons as shown in $10^{15}$eV. We can say the case with \textit{the shortest range} in Fig.10 (or Fig.11) has a strong contrast to that with \textit{the longest range}. The manner of the energy loss in Fig.10 is drastic with two big catastrophic energy losses due to bremsstrahlungs ($\sim 0.8$km and $4.74$km), while that in Fig.13 is very moderate with no catastrophic energy loss. \textit{The shortest range}, $\sim 7.7$ kilometers (Fig.11), is far shorter compared with \textit{the longest range}, $\sim 57.8$ kilometers (Fig.13). It is seen from Fig.10 and Table 1 in the case of \textit{the shortest range} that bremsstrahlung plays a decisive role as the cause of catastrophic energy loss. $87.5$\% of the total energy is lost by $28$ bremsstrahlungs, $11.9$\% by $5760$ direct electron-positron pair productions and $0.623$\% by $40$ photonuclear interactions. In Fig.13, we give the case with \textit{the longest range}. Here, $57.7$\% of the total energy is lost by $46542$ direct electron-positron pair productions, $36.6$\% by $277$ photonuclear interactions and only $5.71$\% by $162$ bremsstrahlungs in the complete absence of catastrophic energy losses. In Fig.12, we give the case with \textit{the average-like range}. Here, $55.8$\% of the total energy is lost by $29321$ direct electron-positron pair productions, $16.8$\% by $118$ bremsstrahlungs and $27.4$\% by $196$ photonuclear interactions, while, in the real averages ($100,000$ samples), $43.9$\% of the total energy is lost by $2.57 \times 10^{4}$ direct electron-positron pair productions, $33.7$\% by $108$ bremsstrahlungs and $22.5$\% by $172$ photonuclear interactions. Thus, it can be concluded that the diversity among muon propagation with the same primary energy should be noticed.

\subsubsection{Average characteristics of high energy muons with the shortest range, the average-like range, the longest range around the average range}
\label{sec:2.4.2}
In Table 2 (a), we give the ratios of energy losses due to respective stochastic processes to total energy loss in the typical ranges (\textit{the average}, \textit{the average-like}, \textit{the shortest} and \textit{the longest}) for $10^{12}$eV, $10^{15}$eV and $10^{18}$eV. It is clear from the Table that, averagely speaking, high energy muons are lost $\sim 50$\% in the direct electron-positron pair production, $\sim 30$\% in bremsstrahlung and $\sim 20$\% in photonuclear interaction. It is clear from Table 2(a) that the muon with \textit{the longest range} loses $\sim 70$ \% in direct electron-positron pair productions, in long chain of electron pairs induced electromagnetic cascade showers with rather smaller energies, while the muon with \textit{the shortest range} loses $\sim 70$\% to $100$\% of their energy in a few number of bremsstrahlungs (catastrophic energy loss).

\begin{table*}[!t]               
\begin{center}
\caption{The ratios of energies transferred from bremsstrahlung, direct electron-positron pair production and photonuclear interaction to the total energy loss (a) and their ratios expressed in respective average values (b).}
\label{tab:ratio}
\scalebox{0.85}[0.9]{
\begin{tabular}{c|c|c|c|c|c|c}
\hline 
             & \multicolumn{2}{c|}{Brems} & \multicolumn{2}{c|}{Direct Pair} & \multicolumn{2}{|c}{Nuclear}\\ 
\hline
$E_{0}=10^{12}eV$ & (a) & (b) & (a) & (b) & (a) & (b) \\
\hline
$<$Average$>$&3.37$\times 10^{-1}$&$1.00 $&5.26$\times 10^{-1}$&$1.00                $&1.37$\times 10^{-1}$&$1.00 $\\ \hline
Average-like &2.87$\times 10^{-1}$&$0.872$&4.30$\times 10^{-1}$&$0.688               $&2.83$\times 10^{-1}$&$2.51 $\\ \hline
Shortest     &5.32$\times 10^{-1}$&$2.96 $&3.76$\times 10^{-1}$&$9.14 \times 10^{-4} $&9.20$\times 10^{-2}$&$2.96 \times 10^{-3} $\\ \hline
Longest      &3.50$\times 10^{-2}$&$0.875$&9.54$\times 10^{-1}$&$1.33                $&1.10$\times 10^{-2}$&$5.87 \times 10^{-2} $\\ \hline 
$E_{0}=10^{15}eV$\\ 
\hline
$<$Average$>$&3.40$\times 10^{-1}$&$1.00 $&4.98$\times 10^{-1}$&$1.00 $&1.62$\times 10^{-1}$&$1.00 $\\ \hline
Average-like &7.54$\times 10^{-1}$&$1.25 $&2.37$\times 10^{-1}$&$0.960$&9.36$\times 10^{-3}$&$0.600 $\\ \hline
Shortest     &8.66$\times 10^{-1}$&$2.53 $&1.34$\times 10^{-1}$&$0.178$&5.90$\times 10^{-5}$&$0.321$\\ \hline
Longest      &7.62$\times 10^{-2}$&$0.541$&8.11$\times 10^{-1}$&$1.52 $&1.12$\times 10^{-1}$&$0.364$\\ \hline
$E_{0}=10^{18}eV$\\ 
\hline
$<$Average$>$&3.24$\times 10^{-1}$&$1.00 $&4.59$\times 10^{-1}$&$1.00 $&2.17$\times 10^{-1}$&$1.00 $\\ \hline
Average-like &1.68$\times 10^{-1}$&$1.30 $&5.58$\times 10^{-1}$&$0.858$&2.74$\times 10^{-1}$&$0.848$\\ \hline
Shortest     &8.75$\times 10^{-1}$&$2.17 $&1.19$\times 10^{-1}$&$0.634$&6.23$\times 10^{-3}$&$2.20 \times 10^{-2}  $\\ \hline
Longest      &5.71$\times 10^{-2}$&$0.209$&5.77$\times 10^{-1}$&$1.49 $&3.66$\times 10^{-1}$&$1.13 $\\ \hline
\end{tabular}
}
\end{center}
\end{table*}

In Table 2(b), the ratios of fractional energy lose for specified stochastic processes are divided by the corresponding averaged ones. It is clear from the Table 2(b) that the divisions of energy loss to the specified processes in \textit{the average-like range} are clearly different from that of the real averaged. It shows that the energy divisions for different processes are different, even if the ranges are same. This fact makes the ejection of the Cherenkov light influence, even if the paths of the high energy muons are same.

In Table 2(b), it is also clear from the characteristics of the typical showers from the point of energy dissipation that energy losses ratios of showers concerned to their averages due to bremsstrahlung in \textit{the shortest ranges} lose their energies are $2.96$, $2.53$, $2.17$ for primaries $10^{12}$eV, $10^{15}$eV and $10^{18}$eV, respectively. Namely, these showers with \textit{the shortest range} essentially lose their energies almost due to bremsstrahlung, while the corresponding ratios due to direct electron-positron pair production in the showers with \textit{the longest range} are $1.33$, $1.52$, $1.49$. Also, these showers with \textit{the longest range} lose their pretty energies owing to direct electron-positron pair production.

\subsubsection{Range Distributions and Hypothetical Range Distributions for high energy muons}
\label{sec:2.4.3}
As shown, for example, in Fig.3 to Fig.13, we can pursue three kinds of the typical types of the behaviors of high energy muons with definite primary energies in stochastic manner exactly, recording the locations of the interaction points for specified interactions and their dissipated energies exactly. However, we pursue the behaviors of all sampled muons exactly, including three different types of the muons. We can construct the range distributions from ensemble of $100,000$ individual muons for respective primary muon's energy, as shown in Fig.14. In the figure, we give $P(R;E_{0})$, the probabilities for the range distribution in water with primary energies, $10^{12}$eV to $10^{15}$eV and $10^{18}$eV in water whose minimum energy is $10^{9}$eV(1GeV), respectively. It is clear from the figure that the width of the range distribution increases rapidly, as their primary energy increases. Also, as the primary energy decreases, the width of range distribution becomes narrower and approaches to a $\delta $ function-type, the limit of which denotes no fluctuation. It is interesting that the range distributions can be well approximated as the normal distribution above $\sim 10^{14}$eV where the total Cherenkov light yields comes almost from the muon induced electromagnetic cascade showers and they are given as,
\begin{figure}[h]
\begin{center}
\resizebox{0.45\textwidth}{!}{\includegraphics{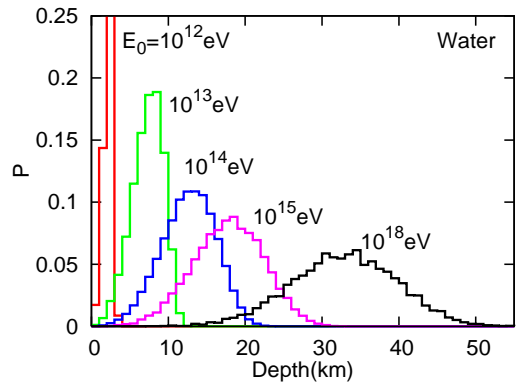}}	   
\caption{Range distributions for $10^{12}$eV to $10^{18}$eV muons in water. The minimum observation energies are taken as $10^{9}$ eV. Each sampling number is 100,000.}
\label{fig:RF131518}
\end{center}
\end{figure}
\begin{equation}              
\label{PRE}                         
P\left(R;E_{0}\right)=\frac{1}{\sqrt{2\pi}\sigma}exp\left(-\frac{R-<R>}{2\sigma^{2}}\right),
\end{equation}
, where $E_{0}$, $R$, $<R>$ and $\sigma $ are primary energy, real range, the average value of ranges and the standard deviations, respectively. Their average ranges, standard deviations and relative variances (standard deviations divided by averages) in water are given in Table 3. Also, it is interesting that their relative variances decrease slightly as their primary energies increase. It should be noticed from Table 3 that the standard deviation increases as primary energy increases, but, the relative variance of the range distribution decreases inversely.
\begin{table}[!t]             
\begin{center}
\caption{The average values, the standard deviations and the relative variances of the range distributions of muons from $10^{11}$eV to $10^{18}$eV in water.}
\label{tab:AVE}
\scalebox{0.9}[0.9]{
\begin{tabular}{c|l|l|c}
\hline
\noalign{\smallskip}
$E_{0}$ [eV] & $<R>$ [km]        & $\sigma$ [km]        & $\sigma/<R>$         \\
\noalign{\smallskip}
\hline
$10^{11}$ & 3.56$\times 10^{-1}$ & 2.52$\times 10^{-2}$ & 7.07$\times 10^{-2}$ \\ \hline
$10^{12}$ & 2.43                 & 4.71$\times 10^{-1}$ & 1.94$\times 10^{-1}$ \\ \hline
$10^{13}$ & 7.28                 & 2.02                 & 2.78$\times 10^{-1}$ \\ \hline
$10^{14}$ & 1.26$\times 10^{1} $ & 3.49                 & 2.77$\times 10^{-1}$ \\ \hline
$10^{15}$ & 1.78$\times 10^{1} $ & 4.57                 & 2.57$\times 10^{-1}$ \\ \hline
$10^{16}$ & 2.30$\times 10^{1} $ & 5.41                 & 2.36$\times 10^{-1}$ \\ \hline
$10^{17}$ & 2.79$\times 10^{1} $ & 6.14                 & 2.20$\times 10^{-1}$ \\ \hline
$10^{18}$ & 3.29$\times 10^{1} $ & 6.81                 & 2.07$\times 10^{-1}$ \\ \hline
\noalign{\smallskip}
\end{tabular}
}
\end{center}
\end{table}
In order to examine each characteristic of the stochastic process, such as the bremsstrahlung, direct electron-positron pair production and photonuclear interaction, we construct the hypothetical range distribution in which a specified stochastic process only is assumed to occur and the other two stochastic processes are assumed not to occur. To clarify the characteristics of the specified stochastic processes, we can compare this hypothetical range distribution with that of real range distribution in which every specified stochastic process is realized as the competition effect among these three processes. We compare the hypothetical range distribution with the real range distribution in Fig.15 to Fig.17.

\begin{figure}[!t]
\begin{center}
\resizebox{0.45\textwidth}{!}{\includegraphics{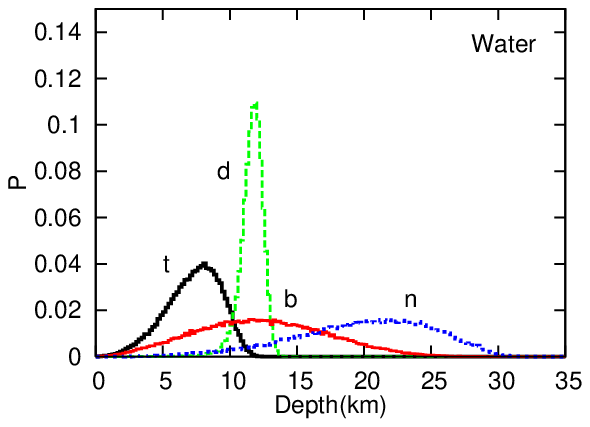}}	       
\caption{Hypothetical range distributions in water for $10^{13}$eV muons together with the real range distribution.}
\label{fig:RF13Sp}
\resizebox{0.45\textwidth}{!}{\includegraphics{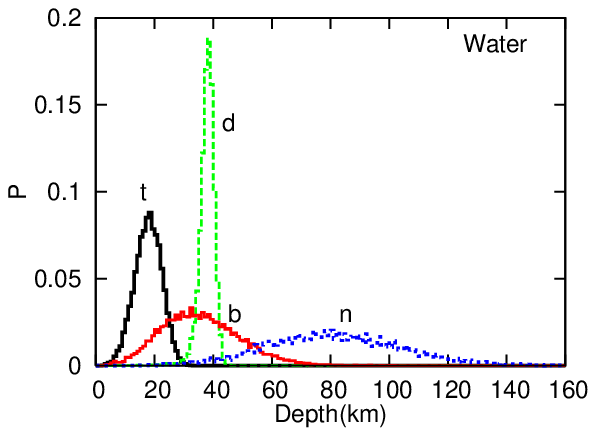}}	       
\caption{Hypothetical range distributions in water for $10^{15}$eV muons together with the real range distribution.}
\label{fig:RF15Sp}
\resizebox{0.45\textwidth}{!}{\includegraphics{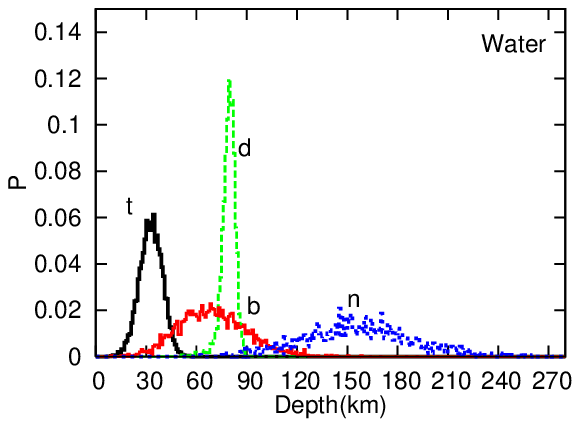}}	       
\caption{Hypothetical range distributions in water for $10^{18}$eV muons together with the real range distribution.}
\label{fig:RF18Sp}
\end{center}
\end{figure}

In Fig.15, we compare three different hypothetical range distributions with the real range distribution for primary energy of $10^{13}$eV. Here, the symbol \textbf{d} in these figures means the hypothetical range distribution in which only direct electron-positron pair production is taken into account and both the bremsstrahlung and photo nuclear interaction are neglected. The symbols \textbf{b} and \textbf{n} have similar meaning to that of \textbf{d}. The symbol \textbf{t} means the real range distribution in which all interactions are taken into account (The true distribution). From the shapes of the distributions and their maximum frequencies for different stochastic processes in the figures, it is clear that energy losses in the direct electron-positron pair production are of small fluctuation, while both the bremsstrahlung and photonuclear interaction are of bigger fluctuation, and the fluctuation in photonuclear interaction becomes bigger when compared with bremsstrahlung as primary energy increases. The smaller fluctuation in direct electron-positron pair production suggests us that energy loss from this process may be treated as something like "continuous" energy loss in the special situation

\subsubsection{Other physical quantities obtained from \textit{the Time Sequential Procedure}}
\label{sec:2.4.4}
In Fig.18 to Fig.20, we give the survival probabilities for different cutoff energies with primary energies of $10^{12}$eV, $10^{15}$eV and $10^{18}$eV, respectively. The values for cutoff energies are given in respective figures. The sampling number utilized is $100,000$ for each primary energy.

\begin{figure}[!t]
\begin{center}
\resizebox{0.45\textwidth}{!}{\includegraphics{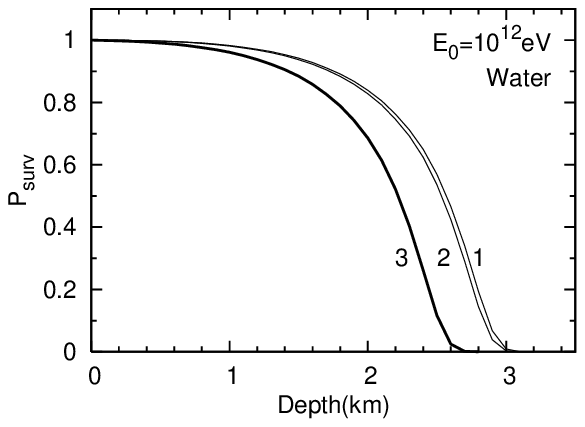}}	        
\caption{The survival probabilities for $10^{12}$eV muon.
Curves labels correspod to following set of cutoff energies:
(1)$10^{9}$eV, (2)$10^{10}$eV, (3)$10^{11}$eV.}
\label{fig:SP12}
\resizebox{0.45\textwidth}{!}{\includegraphics{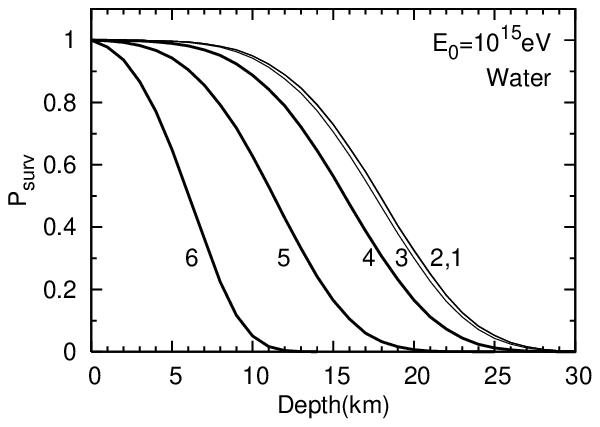}}	        
\caption{The survival probabilities for $10^{15}$eV muon.
Curves labels correspod to following set of cutoff energies:
from (1)$10^{9}$eV to (6)$10^{14}$eV.}
\label{fig:SP15}
\resizebox{0.45\textwidth}{!}{\includegraphics{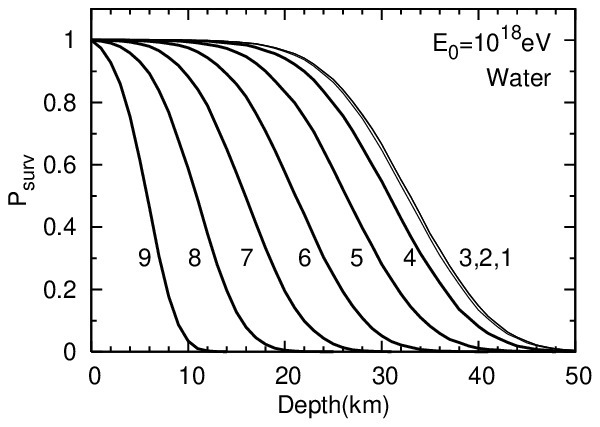}}	        
\caption{The survival probabilities for $10^{18}$eV muon.
Curves labels correspod to following set of cutoff energies:
from (1)$10^{9}$eV to (9)$10^{17}$eV.}
\label{fig:SP18}
\end{center}
\end{figure} 

In Fig.21 to Fig.23, we give the differential energy spectrum of muons for primary energies, $10^{12}$eV, $10^{15}$eV and $10^{18}$eV, respectively. The energy spectra of the survival muons obtained by \textit{the} $V_{cut}$ \textit{Procedure} are surmised to become different from those by \textit{the Time Sequential Procedure}.

As the primary energies of the mouns increase and/or the depths increase, the magnitude of the energy spectra of the muons obtained by \textit{the} $V_{cut}$ \textit{Procedure} is surmised to decrease particularly at lower energies due to the constant $v_{cut}$, compared with those obtained by \textit{the Time Sequential Procedure}. See, further discussion in the next section.

\begin{figure}[!t]
\begin{center}
\resizebox{0.45\textwidth}{!}{\includegraphics{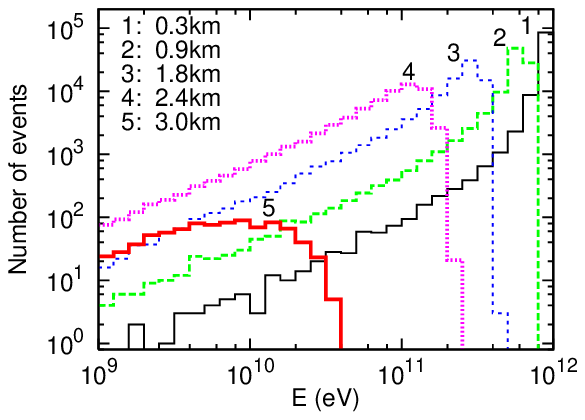}}           
\caption{Energy spectrum in water at the different depths, initiated by $10^{12}$eV muons.}
\label{fig:ES12}
\resizebox{0.45\textwidth}{!}{\includegraphics{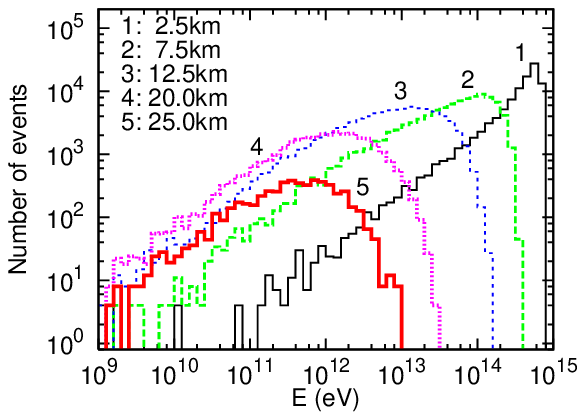}}           
\caption{Energy spectrum in water at the different depths, initiated by $10^{15}$eV muons.}
\label{fig:ES15}
\resizebox{0.45\textwidth}{!}{\includegraphics{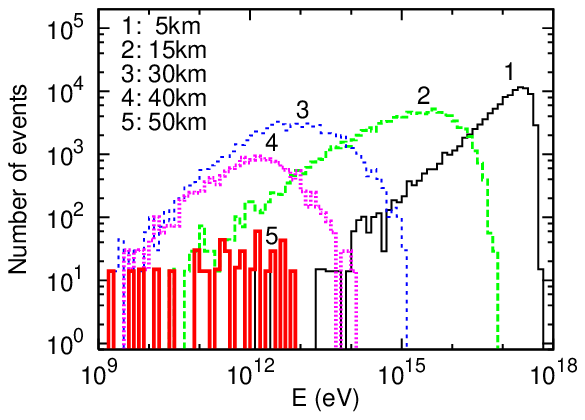}}           
\caption{Energy spectrum in water at the different depths, initiated by $10^{18}$eV muons.}
\label{fig:ES18}
\end{center}
\end{figure}

\section{\textit{The} $V_{cut}$ \textit{Procedure}: The fundamental structure and  its application limit}
\label{sec:3} 

Here, we examine the fundamental structure of \textit{the} $V_{cut}$ \textit{Procedure} from a different point of view.

Lipari and Stanev~\cite{Lipari} and subsequent authors, P.Antonioli, S.Iyer et al., Klimushin et al., D.Chirkin et al., S.Bottai et al.~\cite{Antoni}-~\cite{Chirkin} formulate \textit{the} $V_{cut}$ \textit{Procedure} as follows:

\begin{eqnarray}
\label{dEdx_rad}                
\frac{dE}{dx} &=&\left[\frac{dE}{dx}\right]_{soft}+\left[\frac{dE}{dx}\right]_{hard} \nonumber \\
&=&\frac{N}{A}E\int_{0}^{v_{cut}}dv\cdot v\frac{\sigma\left(v,E\right)}{dv}
\nonumber \\
&+&\frac{N}{A}E\int_{v_{cut}}^{1}dv\cdot v\frac{d\sigma\left(v,E\right)}{dv},
\end{eqnarray}

, where $v$ denotes the fractional emitted energy. They introduce $v_{cut}$ , a certain constant value, into the diffusion equation, in such a way that the effective energy loss, for example, the emitted energies above $v_{cut} \times E$, is treated stochastically in the "hard" part, while that below $v_{cut} \times E$ they are put into "continuous" energy losses (the "soft" part), which are simply subtracted from the muons concerned.

Here, let us summarize the values of $v_{cut}$ utilized in \textit{the} $V_{cut}$ \textit{Procedure} in the following. [a] Lipari and Stanev adopt $v_{cut}$= $0.01$~\cite{Lipari}, [b] Antonioli et al. adopt $v_{cut}$ = $10^{-3}$~\cite{Antoni}, [c] Dutta et al. adopt $v_{cut}$ =$10^{-3}$~\cite{Dutta}, [d] Sokalski et al. adopt $v_{cut}$= $10^{-3}$ to $0.2$~\cite{Sokal}, [e] Chirikin and Rohde adopt $v_{cut}$ = $10^{-4}$ to $10^{-3}$~\cite{Chirkin}.

Relating to its application limit in \textit{the} $V_{cut}$ \textit{Procedure}, the problems to be examined are as follows:

\subsection{The inconsistent treatment in the separation of the "hard" part from the "soft" part}
It should be pointed that the separation of the "soft" part from the "hard" part is treated in inconsistent manner in \textit{the} $V_{cut}$ \textit{Procedure} as for fixed energy muon. Namely, the muon with the some energy is treated in the "soft" part in some case, while the muon with the same energy is treated in the "hard" part in another case. Such the treatment lacks in consistency for description on muon behavior, because the effectiveness of fluctuation depends on the absolute values of muon energies. 

\textit{The} $V_{cut}$ \textit{Procedure} pursues the change of energy state by step by step method with regard to the depth $dx$. Consequently, by the constancy of $v_{cut}$ ($10^{-4}$ under examination), their stochastic energy loss part (the "hard" part) shifts toward lower energy region, as $dx$ advances. In other words, as already mention, the muons with some energy belongs to the "hard" part (stochastic energy loss part) at certain depth, but belongs to the "soft" part ("continuous" energy loss part) at another depth, owing to the shift of the boundary line between the "hard" part and the "soft" part. Such a description on the behavior of the muon in \textit{the} $V_{cut}$ \textit{Procedure} clearly lacks in consistency as for the range fluctuation of high energy muon.

\begin{figure}[h]
\begin{center}
\resizebox{0.45\textwidth}{!}{\includegraphics{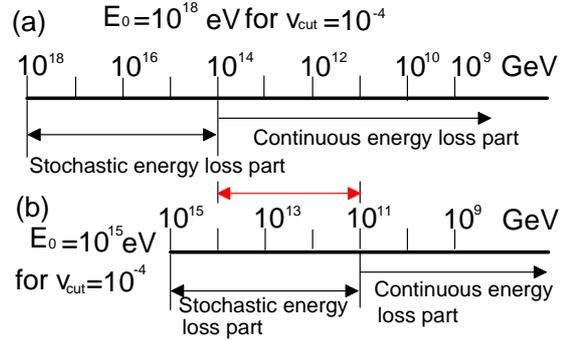}}	   
\caption{(a) The separation of stochastic energy loss part (the "hard" part) from "continuous" energy loss part (the "soft" part) in the case $v_{cut}=10^{-4}$ for $10^{18}$eV. (b) The separation of stochastic energy loss part (the "hard" part) from "continuous" energy loss part (the "soft" part) in the case $v_{cut}=10^{-4}$ for $10^{15}$eV}
\label{fig:Vcut}
\end{center}
\end{figure}

For example, comparing Fig.24(a) with Fig.24(b), it is clear that the region from $10^{14}$eV to $10^{11}$eV for $E_{0}=10^{18}$eV belongs to the "soft" part, while the same region belong to the "hard" part for $E_{0}=10^{15}$eV. This is also an example that the stochastic process is not treated in the unified manner.

The inconsistent description of \textit{the} $V_{cut}$ \textit{Procedure} is clarified by the examination on the interrelation among the $v_{cut}$, the "continuous" energy loss, $E_{b,min}$, the minimum energy of the emitted photon due to bremsstrahlung and $E$, the energy of the muon concerned. In the relation of $v_{cut} = E_{b,min} /E$, there is two choices for given E, namely, which quantities should be chosen as constant (or variable), $v_{cut}$ ? or $E_{b,min}$ ?

In \textit{the} $V_{cut}$ \textit{Procedure}, they adopt $v_{cut}$ to be constant, then, $E_{b,min}$ is the function of $E$. In other word, the values of $E_{b,min}$ change as $E$ change so as to keep $v_{cut}$ to be constant. This denote the borderlines which separate the "soft" part from "hard" part (Eq.(14)) change as $E$ change as shown in Figure 24. Namely, the value of "continuous" energy loss change as the muons concerned change. It should be noted that the "continuous" energy losses are treated as dissipated energies which "flow out" merely from the system for muons towards the outside and neither contribute to the muon propagation any more, nor produce "seeds" for the electromagnetic cascade showers, just as the same in the usual ionization losses. Considering such the character of the "continuous" energy loss, the results that the borderline shift owing to the choice of both $v_{cut}$ and $E$ denote the inconsistent treatment in \textit{the} $V_{cut}$ \textit{Procedure} (see, Figure 24, too). Originally, the borderline should be decided owing to the physical reasons, but it is reluctantly decided in artificial manner.

In Figure 25, the "continuous" energy losses per muon radiation length shown as the function of $v_{cut}$ for given the energies of the muons concerned. Here [\textbf{a1}], [\textbf{a2}], [\textbf{a3}], [\textbf{a4}] and [\textbf{a5}] denote the "continuous" energy loss per muon radiation length ($\sim 1500$ meter in water) for the muons with $10^{18}$ eV, $10^{17}$ eV, $10^{16}$ eV, $10^{15}$ eV and $10^{14}$ eV as the function of $v_{cut}$, respectively. For example, we consider the case of a muon with $10^{16}$ eV ([\textbf{a3}]). In the case of $v_{cut} = 10^{-2}$ [25] and, $10^{-4}$ [31], the "continuous" energy loss per muon radiation length  $\sim 10^{14}$ eV and $\sim 10^{12}$ eV, respectively. Also, we show the energies of the muons concerned where the "continuous" energy losses attain at the usual ionization loss as the function of $v_{cut}$ marked with [c]. It is clear from the figure that the "continuous" energy losses exceed over the usual ionization loss at the energies of the muons concerned for $\sim 3\times 10^{13}$ eV, $\sim 3 \times 10^{14}$ eV, and $\sim 3 \times 10^{15}$ eV for $v_{cut}=10^{-2}$, $10^{-3}$ and $10^{-4}$, respectively, because $E_{b,min}$  increases linearly with $E$. These "continuous" energy losses obtained by \textit{the} $V_{cut}$ \textit{Procedure} are far higher than $\sim 10^{6}$ eV, that obtained by \textit{the Time Sequential Procedure} (see, the section 2.3 and the discussion in the end of this section, too). 

\begin{figure}[t]
\begin{center}
\resizebox{0.45\textwidth}{!}{\includegraphics{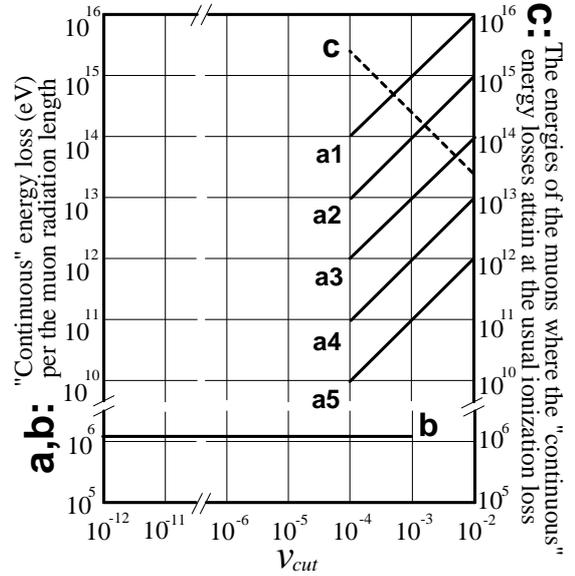}}	         
\caption{The interrelation among $v_{cut}$, the "continuous" energy loss per muon radiation loss in \textit{the} $V_{cut}$ \textit{Procedure} and \textit{the Time Sequential Procedure}, and the "continuous" energy loss at which become equal to the usual ionization loss in \textit{the} $V_{cut}$ \textit{Procedure}.
}
\label{fig:Vcut2}
\end{center}
\end{figure}

It should be noted that the large spread of the numerical values of the "continuous" energy losses obtained by \textit{the} $V_{cut}$ \textit{Procedure} shown in Figure 25 is not owing to the real physical causes, and the introduction of $v_{cut}$ treatment into the muon struggling is made artificially for the sake of convenience to remove the infrared catastrophe in the bremsstrahlung. Namely, the "continuous" energy loss is a kind of artificial product and, therefore, its numerical value is desirable taken to be as small as possible for avoidance of the divergence in bremsstrahlung. Otherwise, the "continuous" energy loss may distort the nature of the original bremsstrahlung. It is easily understood from the figure that \textit{the} $V_{cut}$ \textit{Procedure} is not described in consistent manner owing to the great change in the "continuous" energy loss.

On the other hand, we adopt $E_{b,min}$  to be constant ($1.02$ MeV) in the relation of $v_{cut} = E_{b,min} / E$ in \textit{the Time Sequential Procedure}, instead of $v_{cut}$ to be constant in \textit{the} $V_{cut}$ \textit{Procedure}. Then, $v_{cut}$ is the function of $E$ in our procedure, while $E_{b,min}$ is the function of $E$ in \textit{the} $V_{cut}$ \textit{Procedure}. Owing to the adoption of $E_{b,min}$  to be constant, the "continuous" energy loss per muon radiation unit is $E_{b,min}$  ($1.02$ MeV) by the definition, irrespective of the energies of the muons concerned in \textit{the Time Sequential Procedure} and furthermore, the numerical value $\sim 10^{6}$ eV is far smaller even compared with the usual ionization loss, $\sim 3 \times 10^{11}$ eV. (see, the section 2.3, too). Consequently, the introduction of $v_{cut}$ in \textit{the Time Sequential Procedure} allow us to treat the every energy loss as those really coming from the "radiative" part except for the muons with below $\sim 10^{13}$ eV where we cannot the neglect the effect of the usual ionization loss. Thus, it is conclude that \textit{the Time Sequential Procedure} is described in consistent manner. 

\subsection{From where the differences in the survival probabilities come between \textit{the} $V_{cut}$ \textit{Procedure} and \textit{the Time Sequential Procedure} ?}

For higher energy muons and/or larger $v_{cut}$, for example, in the case of bremsstrahlung, the emitted higher photons may be contained in the "soft" part in which they are subtracted from the muon concerned, being treated "continuous" energy loss and don't contribute to the muon's future behavior any more. However, some part of such the higher photons may not be consumed as dissipated energies,  if $v_{cut}$ is taken up  smaller. 

Then the muons concerned should maintain them in the "hard" part and their energy loss may be treated in stochastic manner. As the result of it, the emitted photons may be correctly taken into account in the "hard" part so that the muon concerned can maintain higher energy than that in the case of larger $v_{cut}$ and, consequently, more muons may survive than in the case of larger $v_{cut}$.

It may be possible to re-state the matter mentioned above in the following.

When one adopts larger $v_{cut}$, one may expect the some deficit in the lower energies above $E_{min}$ in the energy spectrum of the survived muon. This deficits of the lower energies in the energy spectrum of the muons concerned above $E_{min}$ due to larger $v_{cut}$ lead to smaller survival probabilities. The deficits are the another representation which the larger energy loss is put into the "continuous" energy loss. Conclusively speaking, the numerical values in the survival probabilities obtained by \textit{the} $V_{cut}$ \textit{Procedure} are expected to approach to those obtained by \textit{the Time Sequential Procedure} as the ratios of the "soft" part to the "hard" part in \textit{the} $V_{cut}$ \textit{Procedure} decrease, as far as the survival probability is concerned.

Thus, in the case of higher energy muons and/or larger $v_{cut}$, we expect the deformed high energy photons (or electrons) spectrum due to the high energy muons  which, in turn, may result in the deformed Cherenkov light spectra, compared with the corresponding ones which can be obtained by \textit{the Time Sequential Procedure}.

\subsection{The difference between \textit{the} $V_{cut}$ \textit{Procedure} and \textit{the Time Sequential Procedure} in the light of the Monte Carlo method}
\label{sec:3.2}

Apart from the largeness of $v_{cut}$ values in \textit{the} $V_{cut}$ \textit{Procedure}, we discuss the difference in the Monte Carlo method between \textit{the} $V_{cut}$ \textit{Procedure} and \textit{the Time Sequential Procedure}.

On the sampling of energy loss in the radiation part (the "hard" part) in \textit{the} $V_{cut}$ \textit{Procedure}, they utilize the corresponding total cross sections, namely, the sum of the bremsstrahlung, direct electron-positron pair production and photo nuclear interaction, but not utilize the cross section for the respective interaction. As far as one is interested exclusively in the energy loss of the muons concerned, this treatment seems to be reasonable. However, in the case when one is interested in the energy determination of the muons through the Cherenkov light yields due to the muons concerned, one need the detailed information around the respective interaction.

In the treatment of the Cherenkov light yields, they considered only in the fluctuation around the total cross section, while we consider the fluctuation around the respective interaction (see, the discussion in the 4. Conclusion and Outlook)

\subsection{Comparison of the survival probabilities obtained by \textit{the Time Sequential Procedure} with those obtained by \textit{the} $V_{cut}$ \textit{Procedure}}

In the previous section, we examine the inconsistency problems involved in \textit{the} $V_{cut}$ \textit{Procedure}. However, \textit{the} $V_{cut}$ \textit{Procedure} may be useful within their application limit.

\begin{figure}[!t]
\begin{center}
\resizebox{0.45\textwidth}{!}{\includegraphics{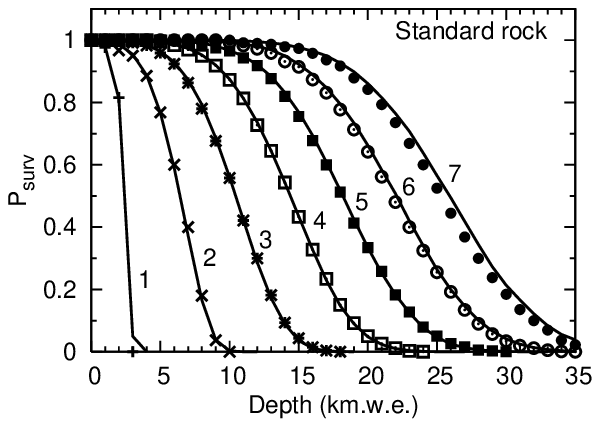}}		 
\caption{The comparison of our result with that of Lipari and Stanev\cite{Lipari}. The survival probabilities of muons of energy from $1$ TeV to $10^{6}$ TeV.  The numerical figures attached each curve denote the primary energies. Curves labels correspond to following set of primay energies of muon: (1)$1$TeV, (2)$10$TeV, (3)$10^{2}$TeV, (4)$10^{3}$TeV, (5)$10^{4}$TeV, (6)$10^{5}$TeV, (7)$10^{6}$TeV. Symbols are due to Lipari and Stanev and curves are due to ours.}
\label{fig:Lipari}
\resizebox{0.45\textwidth}{!}{\includegraphics{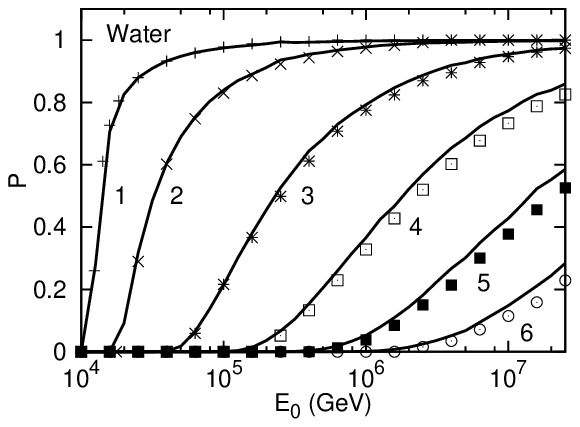}} 
\caption{The comparison of our results with that of Klimushin et al\cite{Klimu}.
The continuous lines are obtained by us, while symbols are readout  from those by Klimushin et al. for primary energies from $10^{13}$ eV to $3\times 10^{16}$ eV. The numerical figures attached each curve denote the threshold energy is 10 TeV. Curves labels correspond to following set of depths: (1)1.15km, (2)3.45km, (3)8.05km, (4)12.65km, (5)17.25km, (6)21.39km.}
\label{fig:Buga}
\end{center}
\end{figure}

In Fig.26 and Fig.27, we give the comparison of our results by \textit{the Time Sequential Procedure} for survival probabilities with those of Lipari and Stanev and those of Klimushin et al. by \textit{the} $V_{cut}$ \textit{Procedure}, respectively. We discuss the agreement or disagreement between the results obtained by \textit{the Time Sequential Procedure} and \textit{the} $V_{cut}$ \textit{Procedure}. The agreement between Lipari and Stanev's (Fig.26) and ours is quite well in the energies from 1 TeV to $10^{4}$TeV, while the disagreement between them become clear beyond $10^{5}$TeV. As indicated in the subsection 3.2 the photons contained the "soft" part, should be involved in the muon spectrum, if the stochastic processes concerned down to $E_{min}$ are taken into account. Therefore, this fact leads increase of survival probability and muon spectrum.
If $v_{cut}$ of Lipari and Stanev in higher primary muon energies is taken smaller value, then, their results are expected to approach to us. The rather nice agreement in lower primary energies between them and us indicates that \textit{the} $V_{cut}$ \textit{Procedure} functions well within its application limit, while the disagreement between them and us in higher primary energies that \textit{the} $V_{cut}$ \textit{Procedure} functions beyond its application limit.

The agreement and disagreement between Klimushin et al's (Fig.27) and ours can be explained similarly. The figure shows that the discrepancies between them and us become larger as the depths become lager. This indicates that the deficit of the lower energies part in survived muon energy spectrum become larger, as the depths increase. Namely, if they adopt smaller $v_{cut}$, their numerical values are expected to approach to us.

Also, it should be noticed that fluctuation effect in the muons' behavior depends entirely on the absolute values of muon's energy.

In \textit{the} $V_{cut}$ \textit{Procedure}, as the Monte Carlo procedure, the calculations are performed by step by step method in $dx$, which inevitably introduce uncertainties due to the accumulation effect in calculation error. One may call their procedure \textit{the differential method}. On the contrast to it, \textit{the Time Sequential Procedure} can determine interaction point directly (\textit{the integral method}). Consequently, its accuracy is independent of the errors due to the accumulation effect coming from $dx$. One may call our procedure as \textit{the integral method} on the contrast to \textit{the differential method}, the relation of which is complementary.

\section{Conclusion and Outlook}
\label{sec:4}
From the methodological point of view, \textit{the Time Sequential Procedure} is classified as \textit{the integral method}, while \textit{the} $V_{cut}$ \textit{Procedures} are done as \textit{the differential method}. They are complementary each other and the results obtained by the present $V_{cut}$ \textit{Procedure} approach to those obtained by \textit{the Time Sequential Procedure}, when both $dx$ and $v_{cut}$ are sufficiently small in the latter. 
We surmise that it takes less time for\textit{ the Time Sequential Procedure} to perform the computation than for \textit{the} $V_{cut}$ \textit{Procedure} to do, when both procedures utilize the same boundary conditions for computation and the results obtained by the both procedures maintain the same accuracy of the computation. 

Main purpose of the development of \textit{the Time Sequential Procedure} is the application to KM3 physics. However, another ones are the application to both the energy spectrum of muons underground or underwater and to the range energy fluctuation problem under different photonuclear interactions (~\cite{Dutta},~\cite{Groom},~\cite{Butk},~\cite{Buga2},~\cite{Petru},~\cite{Kuzm}) because the muon-nucleus inelastic scattering (photonuclear interaction) is of prime interest in another aspect.

In Fig.28, we give the ratios of the Cherenkov light yields due to muon induced electromagnetic cascades showers to the total Cherenkov light ones as the function of the depth traversed for $10^{11}$eV to $10^{16}$ eV. It is clear from the figure that $\sim 10^{11}$eV, the most of the Cherenkov light yield comes from the muon itself, while above $\sim 10^{14}$eV the most Cherenkov light yield comes from the muon induced electromagnetic cascade shower and, consequently, the Cherenkov light yield from the original muon is essentially negligible.

Namely, above $\sim 10^{14}$eV, the Cherenkov light yields are produced essentially from the electromagnetic cascade showers due to the bremsstrahlung, direct electron-positron pair production or photo nuclear interaction.

Now, we discuss this problem in more detail. The electromagnetic cascade showers due to bremsstrahlung, direct electron-positron pair production and photo nuclear interaction are photon induced one, electron pair induced one and aggregate of $\pi_{0}-2\gamma$ induced one, respectively. These different kinds of electromagnetic cascade showers have the different characteristics in their respective behaviors, which show different characteristics in the respective Cherenkov light yields.  In \textit{the Time Sequential Procedure}, we simulate exactly these electromagnetic cascade showers in stochastic manner and calculate the Cherenkov light yields produced from the respective electron segments in the electromagnetic cascade showers concerned (see, Eqs.(5) and (6)).

\begin{figure}[h]
\begin{center}
\resizebox{0.45\textwidth}{!}{\includegraphics{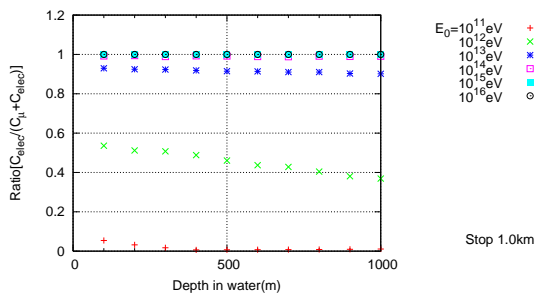}}            
\caption{Ratio of Cherenkov lights due to the accompanied cascade showers to the Cherenkov light.}
\label{fig:RCR}
\end{center}
\end{figure}

In the case that they sample $v$ which is larger than $v_{cut}$ from Eq.(14) in the Monte Carlo simulation of \textit{the} $V_{cut}$ \textit{Procedure}, $v \times E$ denotes the sum of energy losses from bremsstrahlung, direct elctron pair production and photo nuclear interaction with some relative weights, because they are essentially interested in the total energy loss for bremsstrahlung, direct electron-positron pair production and photo nuclear interaction.

Considering the differences in the treatment of the Cherenkov light production through the respective electromagnetic showers among bremsstrahlung, direct electron-positron pair production and photo nuclear interaction in addition to adoption of larger $v_{cut}$ value, the difference in the Cherenkov light production spectrum between \textit{the} $V_{cut}$ \textit{Procedure} and \textit{the Time Sequential Procedure} will become apparent, which is just the main subject in our subsequent papers. A part of the correlations between the energy losses of the high energy muons and their Cherenkov light yields was reported elsewhere~\cite{Okumura}.

However, as far as one is interested in the muons behavior exclusively, the situation mentioned above may be not influential. This is the reason why there are no significant difference in the survival probabilities between \textit{the} $V_{cut}$ \textit{Procedure} and \textit{the Time Sequential Procedure} (see, Figures 26 and 27).

Up to now, our discussion still remains to single high energy muon problem. However, really, muon does not exist singly, but they exist in the form of energy spectrum which is directly reflection of parent neutrino spectrum through the interactions. 

In \textit{the Time Sequential Procedure}, for given primary muon, we exactly simulate the muon behaviors in stochastic manner, without introducing the "soft" part and, consequently, we obtain accurate muon energy spectrum at arbitrary depths as well as the energy spectrum of 'primary' (in the sense of origin of electromagnetic cascade showers) electrons and photons energy spectrum due to the muons concerned at arbitrary depths. The existence of the energy spectrum of primary muons brings more difficulty into the elucidation of fluctuation effect even in \textit{the Time Sequential Procedure}. These effects intermingle with each other and we could not discuss them separately. 

Up to now, we restrict our discussion around electromagnetic cascade showers which the primary muons produce in the case of Bethe-Heitler shower. However, we could not neglect LPM effects, related to the interpretation of extremely high energy muon events in future. One is related to electrons~\cite{Koni1}-\cite{Misaki4} and other is related to muon~\cite{Poly1,Poly2}. We cannot neglect the LPM effect on the behaviors of electromagnetic cascade themselves above $\sim 10^{15}$eV in water~\cite{Misaki5}. However, the LPM effect is supposed to be effective above $\sim 10^{18}$eV in the case of muon induced electromagnetic cascade showers

In such extremely high energies, range fluctuation of muon may alter their feature essentially compared with that of present situation. Furthermore, above $10^{21}$eV, we cannot neglect the LPM effect related to the muons~\cite{Poly1,Poly2}. Namely, above $10^{21}$eV, the Cherenkov light spectrum is supposed to become essentially different from those in the absence of two kinds of LPM effects at present.

In the present paper, we restrict our discussion to muons themselves in high energies. In subsequent papers, we will extend our examination to the Cherenkov light yield via electromagnetic cascade showers from different interactions, such as, bremsstrahlung, direct electron-positron pair production and photonuclear interaction, taking into account of muon energy spectrum for imaging KM3 detector. There, the main subject will be the examination of the Cherenkov light yeilds from the muon induced electromagnetic cascade showers.

In discussion of the problems around high energy neutrino spectrum from the universe, it should be noticed that the reliable results are obtained essentially through the utilization of the stochastically correct tools, taking into account of very few number of experimental events and the steepness of high energy neutrino spectrum.

\section*{Acknowledgments}
One of authors (A.M.) would like to express his thanks to The Institute for China-Japan Culture Study for providing the research fund and for stimulating him.

\bibliography{<your-bib-database>}

\begin{thebibliography}{00}
\bibitem{Menon}         
M.G.K.Menon, P.V.Ramana Murthy, Progress in Elementary Particles and Cosmic Ray Physics, Vol.9 (1967) 161-243, North- Holland Publ.
\bibitem{Miyaz}         
Y.Miyazaki, Phys. Rev. \textbf{76} (1949) 1733
\bibitem{Miyake1}       
S.Miyake, V.S.Narashimham, P.V.Ramana Murthy, Nouvo Cimento \textbf{32} (1964) 1505
\bibitem{Buga}          
E.V.Bugaev, A.Misaki, V.A.Naumov, T.S. Sinegovskaya, S.I.Sinegovsky, N.Takahashi, Phys. Rev. D \textbf{58} (1998) 054001
\bibitem{Mand1}         
M.Mand, L.Ronchi, Nuovo Cimento \textbf{9} (1952) 105
\bibitem{Mand2}         
M.Mand, L.Ronchi, Nuovo Cimento \textbf{9} (1952) 517
\bibitem{Mand3}         
M.Mando, P.G.Sona, Nuovo Cimento 10 (1953) 1275
\bibitem{Rozen}         
I.L.Rozental, V.N.Streltsov, Zh.Eksp. Teor. Fiz \textbf{35} (1958) 1440
\bibitem{Zatse}         
G.T.Zatsepin, E.D.Mikhalchi, Proc. 0th Int. Cosmic Ray Conf., Kyoto, Japan \textbf{3} (1961) 356
\bibitem{Nishi}         
J.Nishimura, Proc. 8th Int. Cosmic Ray Conf., Jaipur, India \textbf{6} (1963) 224
\bibitem{Guren1}        
V.I.Gurentsov, G.T.Zatsepin, E.D.Mikhalchi, Sov.Jour.Nuc.Phys. \textbf{23} (1976) 527
\bibitem{Kobay}         
K.Kobayakawa, Nuovo Cimento B \textbf{47} (1967) 156
\bibitem{Misaki1}       
A.Misaki, J.Nishimura, Uchusen Kenkyuu \textbf{21} (1976) 250; ICR-Report-45774-4, University of Tokyo (1977)
\bibitem{Guren2}        
V.I.Gurentsov, G.T.Zatsepin and E.D. Mikhalchi, Sov.J.Nucl.Phys. \textbf{23} (1976) 527
\bibitem{Minori}        
Y.Minorikawa,T.Kitamura, K.Kobayakawa, Nuovo Cimento C \textbf{4} (1981) 471
\bibitem{Naumov}        
V.A.Naumov, S.I.Sinegovsky, E.V.Bugaev, Phys.Atom.Nucl. 57 (1994) 412
\bibitem{Oda}           
H.Oda, T.Murayama, Jur.Phys.Sos.Japan \textbf{20} (1965) 1549
\bibitem{Bolin}         
H.J.Bolinger, Ph.D thesis, Cornel University (1958)
\bibitem{Hayman1}       
P.J.Hayman, A.W.Wolfendale, Proc. Phys. Soc. \textbf{80} (1962) 710
\bibitem{Hayman2}       
P.J.Hayman, N.S.Palmer, A.W.Wolfendale, Proc. Roy. Soc. A \textbf{275} (1963) 391
\bibitem{Ramana}       
P.V.Ramana Murthy, Ph.D thesis, Bombay University (1962)
\bibitem{Miyake2}       
S.Miyake, V.S.Narasimham, P.V.Ramana Murthy, Nuovo Cimento \textbf{32} (1964) 1524
\bibitem{Taka1}         
N.Takahashi, A.Misaki, A.Adachi, N.Ogita, Y.Okamoto, K.Mitsui, H.Kujirai, S.Miono, O.Saavedra, Proc.18th Int. Cosmic Ray Conf., Bangalore, India \textbf{11} (1983) 443
\bibitem{Taka2}         
N.Takahashi, H.Kujirai, A.Adachi, N.Ogita, A.Misaki, Uchusen Kenkyuu \textbf{28} (1984) 120
\bibitem{Lipari}        
P.Lipari, T.Stanev, Phys. Rev. D \textbf{44} (1991) 3543
\bibitem{Antoni}        
P.Antonioli, C.Ghetti, E.V.Korolkova, V.A. Kudryavtsev, G.Sartorelli, Asroparticle Physics \textbf{7 }(1997) 357
\bibitem{Bottai}        
S.Bottai, L.Perrone, Nucl.Instrment and Method in Research A \textbf{459} (2001) 319
\bibitem{Dutta}         
S.Iyer Dutta, M.H.Reno, I.Sarcevic, D. Seckel, Phys.Rev. D \textbf{63} (2001) 094020
\bibitem{Klimu}         
S.I.Klimushin, E.V.Bugaev, I.A.Sokalski, Phys. Rev. D \textbf{64} (2001) 014016
\bibitem{Sokal}         
L.A.Sokalski, E.V.Bugaev, S.I.Klimushin, Phys. Rev. D \textbf{64} (2001) 074015
\bibitem{Chirkin}       
D.Chirkin, W.Rhode, hep-ph 0407075 v2 (2008)
\bibitem{Kelner}        
For example, S.R.Kelner, R.P.Kokoulin, A.A. Petrukhin, Preprint MEPHI 024-95 Moscow (1995); CERN SCAN-9510048
\bibitem{Kokoulin}      
For example, R.P.Kokoulin, A.A.Petrukhin, Proc. 11st Int. Cosmic Ray Conf., Budapest, Hungary MU-\textbf{41} (1969) 277
\bibitem{Borg}          
For example, V.V.Borog, A.A.Petrukhin, Proc. 14th Int. Cosmic Ray Conf., Munchen, Germany \textbf{6} (1975) 1949
\bibitem{Groom}         
D.E.Groom, N.V.Mokhov, S.I.Striganov, Atomic Data and Nuclear Data Table, \textbf{78} (2001) 183
\bibitem{Tamura}        
M.Tamura, Prog.Theor.Phys. \textbf{34} (1965) 912
\bibitem{Adachi}        
A.Adachi, Y.Fujimoto, N.Ogita, S.Takagi, A.Ueda, Suppl.Prog.Theor.Phys. \textbf{32} (1964) 154
\bibitem{Butk}          
A.V.Butkevich, S.P.Mikheyev, JETP \textbf{95} (2002) 11
\bibitem{Buga2}         
E.V.Bugaev, Yu.V.Shelpin, Phys.Rev.\textbf{D} \textbf{67} (2003) 034027
\bibitem{Petru}         
A.A.Petrukhin, D.A.Timashkov, Phys.At.Nucl. \textbf{67} (2004) 2216


\bibitem{Kuzm} 
K.S.Kuzmin, K.S.Lokhtin, S.I.Sinegovskii, Phys. Part. Nucl. Lett. \textbf{4} (2007) 477
\\
\nonumber
K.S.Kuzmin, K.S.Lokhtin, S.I.Sinegovskii, Int. J. Mod. Phys. A \textbf{20} (2005) 6956

\bibitem{Rossi}         
B.Rossi, K.Greisen, Rev.Mod.Phys. \textbf{13} (1941) 240
\bibitem{Nishi2}        
J.Nishimura, Handbuch der Physik, \textbf{XLVI/2}, 'COSMIC RAY II', Springer-Verlag, (1966) 1
\bibitem{Misaki2}       
A.Misaki, Suppl.Prog.Theor.Phys., Suppl. Theor.Phys. \textbf{89} (1965) 82
\bibitem{Koni1}         
Konishi, A.Misaki, N.Fujimaki, Nuovo Cimento, \textbf{44} A (1978) 509
\bibitem{Stanev}        
T.Stanev, Ch.Vankov, R.E.Streitmatter, R.W. Ellsworth, T.Bowen, Phys. Rev. D \textbf{25} (1982) 1291
\bibitem{Koni2}         
E.Konishi, A.Adachi, N.Takahashi, A.Misaki, J.Phys.G:Nuc.and Part.Phys., 17 (1991)719
\bibitem{Misaki5}       
A.Misaki, Fort.Schr. d. Phys. \textbf{38} (1990) 413
\bibitem{Misaki3}       
A.Misaki, Nuov.Cim. \textbf{13} C (1990) 733
\bibitem{Misaki4}       
A.Misaki, Phys. Rev. D  \textbf{40} (1990) 3086
\bibitem{Poly1}         
S.Polytiko, M.Kato, E.Konishi, N.Takahashi, A.Misaki, J.Phys.G:Nucl.and Part.Phys. \textbf{28} (2002) 427
\bibitem{Poly2}         
S.Polytiko, N.Takahashi, M.Kato, Y.Yamada, A.Misaki, Nucl. Instr. Meth. in Physical Research B \textbf{173} (2001) 30
\bibitem{Okumura}       
Y.Okumura, N.Takahashi, A.Misaki, arXiv: 1010.5054v1 [astro-ph.HE] (2010)
\end{thebibliography}

\end{document}